\documentclass[aps,prb]{revtex4}
\usepackage{bm}
\usepackage{amsmath}
\usepackage{color}
\usepackage{graphicx}
\usepackage{amssymb}
\DeclareMathAlphabet{\mathbbmsl}{U}{bbm}{m}{sl}
\makeatletter
\newsavebox{\@brx}
\newcommand{\llangle}[1][]{\savebox{\@brx}{\(\m@th{#1\langle}\)}%
	\mathopen{\copy\@brx\kern-0.5\wd\@brx\usebox{\@brx}}}
\newcommand{\rrangle}[1][]{\savebox{\@brx}{\(\m@th{#1\rangle}\)}%
	\mathclose{\copy\@brx\kern-0.5\wd\@brx\usebox{\@brx}}}
\makeatother

\begin{document}
\draft


\draft

\title{Tuning excitons and superfluidity of dipolar excitons in the double layers of kagome lattice  by applying circularly polarized irradiation}

\author{ Sita Kandel$^{1,2}$, Godfrey Gumbs$^{1,2,3}$, Teresa Lee$^{1,2}$, and Oleg L. Berman$^{2,4}$  }
\address{$^1$Department of Physics, Hunter College, City University of New York, 695 Park Avenue, New York, NY 10065 USA}
\address{$^{2}$The Graduate School and University Center, The City University of New York, New York, NY 10016, USA}
\address{$^3$Donostia International Physics Center (DIPC), P de Manuel Lardizabal, 4, 20018 San Sebastian, Basque Country, Spain}
\address{$^4$Physics Department, New York City College of Technology, The City University of New York, \\
300 Jay Street,  Brooklyn, NY 11201, USA}

\date{\today}

\begin{abstract}
We present  detailed calculations for several significant properties of the kagome lattice in the presence of irradiation.  We employ the Floquet-Magnus perturbation expansion to  obtain the energy bands and the corresponding wave functions near the Dirac points for the kagome lattice in the presence of circularly or linearly polarized irradiation. In  contrast with linearly polarized irradiation for which the energy bands do not get modified, a band gap is opened up near the Dirac points, between the valence and conduction bands in the presence of circularly polarized irradiation.  We calculated the exciton binding energy, and the exciton energy   for gapped kagome lattice as a function of the frequency and intensity of the irradiation.  We compare the exciton binding energy and exciton energy in a monolayer with those in a double layer consisting of electrons in one layer and holes in a parallel layer, separated by an insulator to inhibit recombination. We predict that a phase transition in the kagome lattice from the semiconducting phase to the excitonic insulating  phase can be induced by applying circularly polarized irradiation. We examined the conditions for  such a phase  transition.  Superfluidity of dipolar excitons  was investigated  as well as the collective properties of the kagome lattice  by calculating the sum of ladder diagrams for the vertex, describing the dipole-dipole repulsion between excitons.   We propose observation of  Bose-Einstein condensation and superfluidity of quasi-two-dimensional dipolar excitons  in two-layer kagome lattices in the presence of pumping by circularly polarized light. The energy spectrum of collective excitations and the sound velocity, as well as the effective mass of dipolar excitons are obtained in the regime of pumping by circularly polarized light. The superfluid density $n_{s}$ and the temperature of the Kosterlitz-Thouless phase transition $T_{c}$ are shown to be monotonic increasing functions of the excitonic density $n$ and the interlayer separation $D$. We have also analyzed the dependence of superfluid density $n_{s}$ and the temperature of the Kosterlitz-Thouless phase transition temperature on the parameters for circularly polarized light. We explore opportunities to tune exciton binding energy, the spectrum of collective excitations, and the critical temperature of the superfluidity by applying circularly polarized irradiation.
\end{abstract}

\maketitle

\medskip

\medskip

\noindent
{\bf  Corresponding author}:\ \     Sita Kandel;   E-mail:   skandel@gradcenter.cuny.edu

\section{Introduction}
\label{sec1}

The rare-earth metals  are a group of inter-metallic elements nearly similar in their appearance - silvery-white and soft. These compounds contain heavy metals and have a variety of applications including lasers, magnetic materials, electrical and electronic components, amorphous materials and glass, as well as industrial processes.  Several heavy-fermion metals having lattices  with geometrical frustration  have revealed  unconventional metallic behavior. \cite{KS}     These ``complex structures” include  the    kagome CeRhSn  \cite{K1,K2}, CePdAl \cite{K3,K4,K5}, YbAgGe, \cite{K6,K7}    YbPdAs \cite{K8}, and pyrochlore Pr2Ir2O7    \cite{K9,K10,K11}.   It has been demonstrated recently  that  Ni$_3$In is a correlated kagome metal \cite{K12}, and it has been suggested that so is Ce$_3$Al  \cite{K13}.

\medskip
\par

In a recent paper, the authors \cite{Naf2} provided a convincing argument of how band topology in kagome systems can be influenced by spin-orbit interactions and electron correlations. These could be relevant to exploring Schrödinger-like corrections to the Dirac equation. 
The way  in which the kagome lattice band structure is affected by strain was investigated in Ref.  \cite{Naf1}.  It was demonstrated that strain leads to  shifting Dirac points and modification of the flat band. Under uniaxial strain, the system sustains an anisotropic energy dispersion relation where one direction remains nearly dispersionless whereas the other acquires a quadratic component. This seems to directly connect to the idea of a hybrid Dirac-Schr\"{o}dinger equation.
\medskip
\par

\begin{figure} 
\centering
\includegraphics[width=0.6\textwidth]{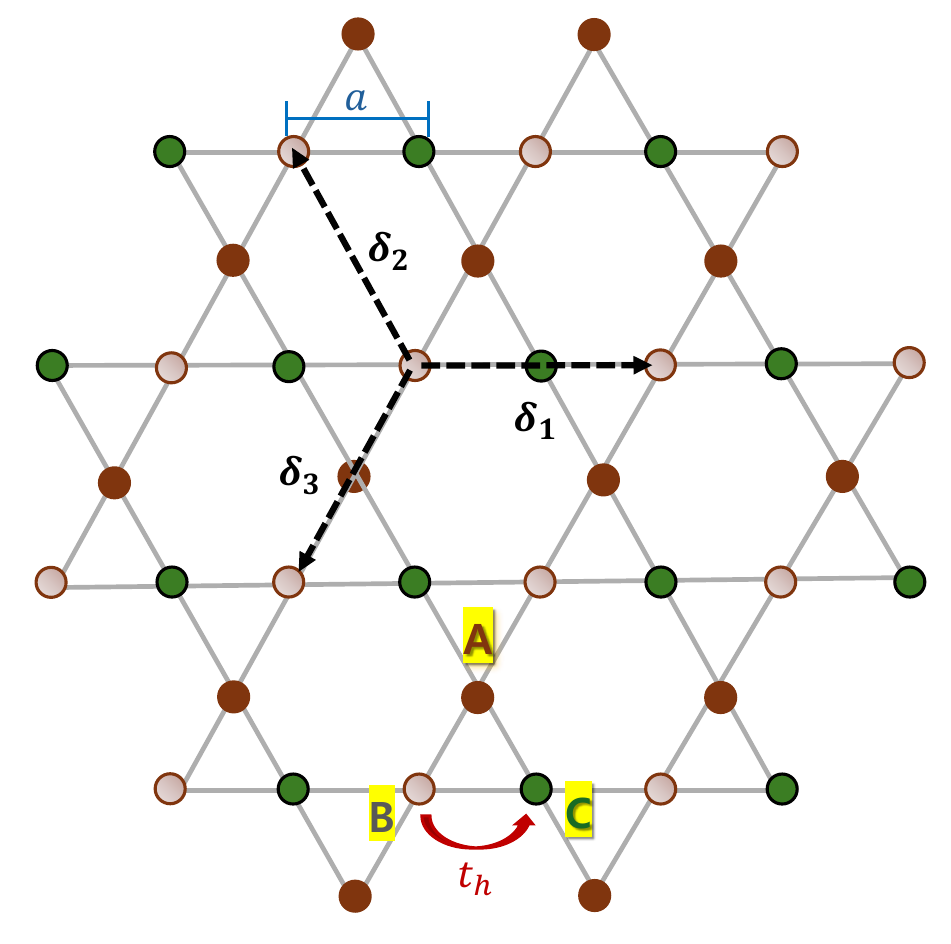}
\caption{(Color online)     Schematic illustration of a kagome lattice. The letters A, B, and C represent the three sublattices. The blue line indicates the nearest-neighbor separation $a$. The red arrow indicates the nearest-neighbor hopping $t_{h}$ while the dotted arrows represent the primitive lattice translation vectors $\boldsymbol\delta_{1}$, $\boldsymbol\delta_{2}$, and $\boldsymbol\delta_{3}$. }
\label{FIG:0}
\end{figure}

The kagome lattice  is triangular with a basis of three lattice points.  See Fig. \ref{FIG:0}. This unusual arrangement  makes the kagome lattice ideal for studying geometric frustration  \cite{lacroix2011} and the resulting exotic quantum states of matter known as quantum spin liquids \cite{lee2007, balents2010, yan2011, jiang2012, han2012, norman2016, liao2017, zhou2017, lauchli2019}. Additionally, from a structural perspective, the wave function associated with a hexagonal ring in the kagome lattice is localized due to the destructive interference for the wave functions at the corner sites \cite{bergman2008,yang2014}. This results in highly degenerate dispersionless bands \cite{mielke1991, mielke1992}  which are stable against disorder   \cite{bilitewski2018}. Using the tight-binding model, a detailed investigation of the band structure for a wide variety of honeycomb and kagome systems has been reported in Ref.  [\onlinecite{mallah}]. Some solid state systems with itinerant electrons also behave like a kagome lattice  as reported in Refs. [\onlinecite{Mazin2014,Ye2018, Leykam2018}].

\medskip
\par

The electronic band structure of kagome lattices typically consists of a Dirac-like band plus a flat band, both  of which can be generated with the use of the tight-binding method. Furthermore,  
single-orbital nearest-neighbor hopping can be employed in these calculations.  The topologically protected linearly dispersive bands have different behaviors from the traditional parabolic bands.  A small band gap can open when spin-orbit coupling (SOC) is taken into account.   In this way,  a mass term can be added to the linearly dispersive bands and a massive Dirac fermion thus can be formed.  The light band (Dirac fermions) coexist with the extremely heavy band (flat bands) in kagome metals. There would be unusual phenomena occurring if either of these bands is tuned close to the Fermi level.  Due to the similarity which  flat bands  have with Landau levels in 2D  materials, flat bands can be responsible for novel quantum physics.  These include magnetism, fractional quantum Hall states, and Mott insulators, and superconductivity \cite{nature18, nature19,nature20},  as it was predicted for twisted bilayer graphene \cite{nature21,nature22}.

\medskip
\par

We want to make it clear that in this work, we confine our attention to a finite range of momentum. Here, we do not include in our investigation the super-heavy localized electrons/quasiparticles in the dispersionless flat band.  In contrast,  we focus on the linear band hosting massless or lightweight quasiparticles. Our major objective is to open a gap at the Dirac point with Floquet engineering using polarized light irradiation. 
 
\medskip
\par

We consider two parallel kagome layers separated by a slab of insulating material  subjected to  pumping by circularly polarized light. The equilibrium system of local pairs of spatially separated electrons and holes can be created by varying the chemical potential with the use of a bias voltage between two kagome layers or between two gates located near the corresponding  kagome lattice sheets (for simplicity, we also call these equilibrium local e-h pairs as dipolar excitons).   Excitons with spatially separated electrons and holes can also be created by laser pumping and by applying a perpendicular electric field as for coupled quantum wells (CQWs). We assume the system is in a quasi-equilibrium
state. Below, we assume a low-density regime for excitons, i.e.,  the exciton radius $r_0 \ll  n^{-1/2}$, where $n$ is the 2D exciton density.

\medskip
\par
Systems of dipolar, or spatially indirect excitons, and bound states of spatially separated electrons and holes, were discussed  in several  theoretical and experimental studies. These structures are known to be formed in coupled quantum wells (CQWs) in semiconductors, or in double layers of two-dimensional  (2D) anomaterials. These systems  can exhibit Bose-Einstein condensation (BEC) and superfluidity
\cite{para1,para2,littlewood,para4,para5,para6}. A detailed review on experimental and theoretical works on the superfluidity of  dipolar excitons was performed in Ref. \cite{para7}. Such superfluidity can occur in the BEC regime, and also in the
Bardeen-Cooper-Schrieffer (BCS) crossover regime \cite{para8}.

It has been predicted that many different systems of dipolar
excitons in double layers are expected to exhibit a superfluid phase
\cite{littlewood,para19,para16,para17,dipolar1,dipolar2,dipolar3,dipolar4}.
This is due to the fact that the energy dispersion of the collective
excitations in a weakly interacting Bose gas of dipolar excitons
satisfies the Landau's criterion for superfluidity
\cite{landau,abrikosov,pitbook}. Superfluidity of excitons in an
\textit{h}-BN-separated MoSe$_2$/WSe$_2$ heterostructure has been
recently observed \cite{expsec2}.

In this paper, we will demonstrate the  occurrence of superfluidity
of  dipolar excitons in the double layers of kagome lattices. We
obtain the exciton binding energy, the spectrum of collective
excitations, the concentration of the superfluid component and the
critical temperature of the superfluidity as functions of the
circularly polarized radiation. We explore opportunities to tune
these properties by the circularly polarized radiation. In addition,
we propose controlled by the circularly polarized radiation the
phase transition from the semiconductor phase to the excitonic
insulator phase (EI) in kagome lattices.

\medskip
\par
The organization of the remainder of this paper is as follows. In Sec. \ref{sec2},  we present the kagome model Hamiltonian and the associated energy band structure which was obtained using the tight-binding approach.  There are Dirac points as well as a flat band.  We theoretically investigate the evolution of  the electronic band structure of the kagome lattices in the vicinity of the Dirac point crossings in response to linearly and circularly polarized irradiation.  These electron dressed states  are achieved in Sec. \ref{sec3} with the aid of the  high-frequency Floquet-Magnus perturbation expansion. We obtain an analytic expression for the energy gap  between the valence band and conduction band. This gap is a function of the	 frequency and electric field of the electromagnetic field and is therefore tunable.   Additionally, although the effective masses of the electron and hole near the gap are isotropic, their group velocities are not, as  presented in Eq.\ (\ref{E_eff_C2}),  thereby making the kagome  materials different from graphene with Dirac cones. In Sec. \ref{sec4},  we consider interacting electron-hole pairs in a monolayer with a kagome lattice. This results  in a two dimensional hydrogen-like atom.  We calculate the exciton binding energy (a bound state of an electron-hole pair) in the presence of circularly polarized light irradiation. We then turn our attention  in Sec. \ref{sec5}  to dipolar excitons in double layer kagome lattices. There, electrons are confined to one layer and holes in the other, with an insulator between them so as to inhibit recombination. We calculate the exciton binding energy and obtain analytical results for small and wide separations. Sec. \ref{sec6}   is devoted to collective excitations  for spatially separated layers of electrons and holes.  In Sec. \ref{sec7}, we present tunable superfluidity of dipolar excitons in double layers of kagome lattices. In Sec. \ref{sec8},  we conclude with a summary of our new results and a discussion.

\section{Kagome Model}
\label{sec2}

\begin{figure} 
\centering
\includegraphics[width=0.8\textwidth]{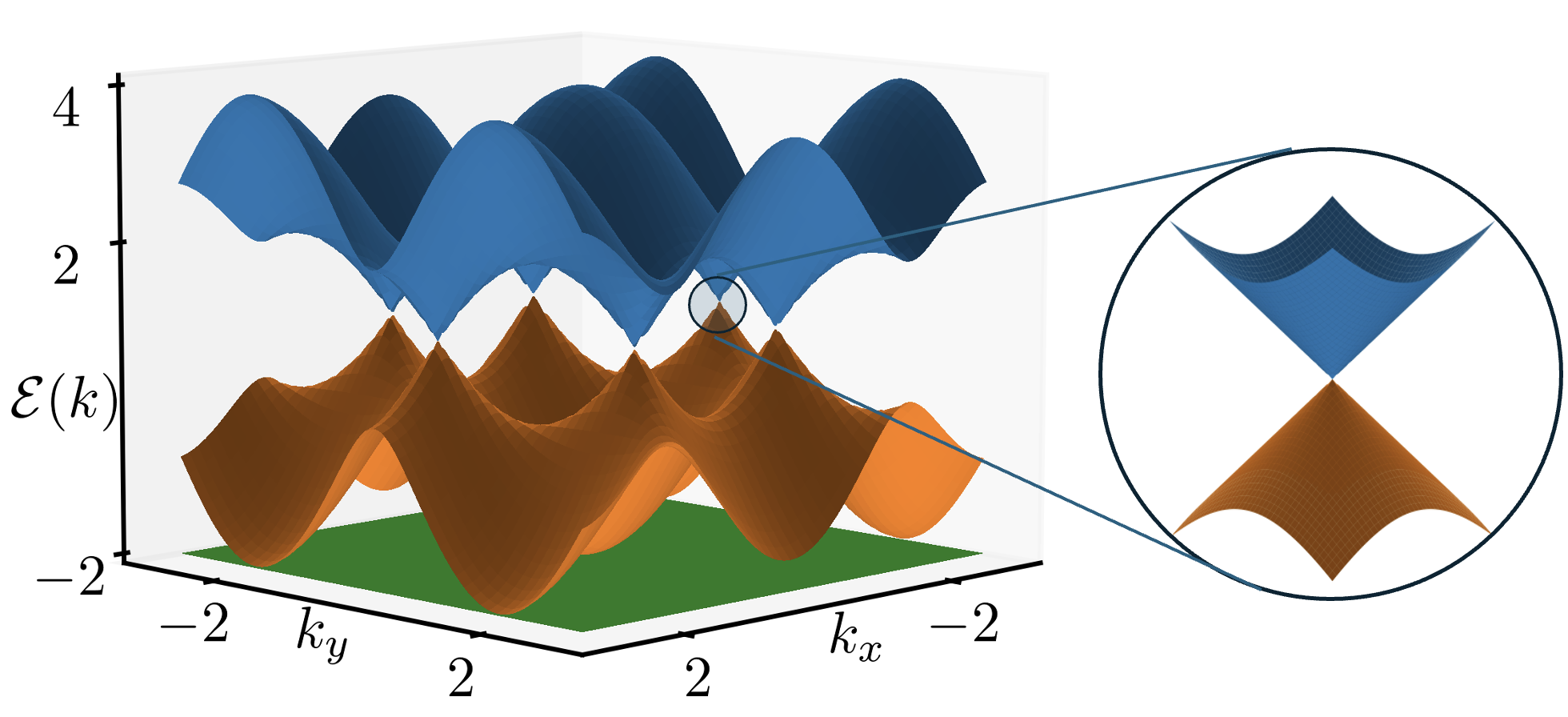}
\caption{(Color online) 
Three dimensional representation of the band structure of the kagome lattice in units of the hopping parameter $t_h$. It shows two dispersive bands, one conduction and one valance, and another which is flat located at -2$t_h$. The bands are plotted as  functions of the wavevector ${\bf k} = (k_x, k_y)$, measured from the $\Gamma$ point at the center of the Brillouin zone. }
\label{FIG:1}
\end{figure}

The kagome lattice has a non-Bravais lattice with a unit cell containing three atomic sites as depicted in Fig. \ref{FIG:0}. The locations of these three atoms in the unit cell are expressed in terms of the nearest-neighbor separation ``a”. The tight-binding Hamiltonian for the kagome lattice  with hopping parameter $t_h$ is given by \cite{Eugene}

\begin{equation}
\hat{H} =  t_{h} \sum_{<i, j, l> \sigma} \left(\hat{a}^{\dag}_{i,\sigma}\hat{b}_{j,\sigma}+ \hat{a}^{\dag}_{i ,\sigma}\hat{c}_{l,\sigma} +\hat{b}^{\dag}_{j,\sigma}\hat{c}_{l,\sigma} + h.c.\right)\  ,  
\label{K_Hamil_Tb}
\end{equation}
where $\hat{a}, \hat{b}$ and $\hat{c}$ are quantum operators acting respectively on the three different sublattices. The indices $i, j, l$ refer to the real space positions of the lattice sites and $\sigma$ is a spin label. The Hamiltonian  in  Fourier space can be expressed in the form of a $3\times3$ matrix which is 

\begin{equation}
\hat{H}({\bf k})   =  t_h \  \left(
 \begin{array}{ccc}
 0 &1+ e^{-i\bf {k} .  \boldsymbol\delta_1} &  1+ e^{i\bf {k} .  \boldsymbol\delta_3} \\
   1+ e^{i\bf {k} .  \boldsymbol\delta_1} & 0 &  1+ e^{-i\bf {k} .  \boldsymbol\delta_2}  \\
     1+ e^{-i\bf {k} .  \boldsymbol\delta_3} &1+ e^{i\bf {k} .  \boldsymbol\delta_2}& 0  
 \end{array}
 \right) ,
 \label{K_Hamil_F}
\end{equation}
where ${\bf k}$ is a wave vector measured from the $\Gamma$ point at the center of the Brillouin zone of the reciprocal lattice and $\boldsymbol\delta_1 = (2, 0) a, \boldsymbol\delta_2 = (-1, \sqrt{3}) a, \boldsymbol\delta_3 = (-1, -\sqrt{3}) a $ are the primitive lattice vectors in units of the nearest-neighbor distance $a $. On diagonalization of this matrix Hamiltonian, we obtain three energy bands of the kagome lattice as 

\begin{eqnarray}
  \label{eigen_value2}
  {\cal E}_s({\bf k}) =\left\{
 \begin{array}{cc}
 -2t_h & for \, s = 0 \\ \\
 t_h\left(1+s \sqrt{3+2 \sum_{n} \cos(\bf {k} .  \boldsymbol\delta_n)}\,  \right)& for \, s = \pm 1
 \end{array}%
 \right. 
\end{eqnarray}
with $\boldsymbol\delta_n = (\boldsymbol\delta_1, \boldsymbol\delta_2, \boldsymbol\delta_3)$. One of these energy bands is a dispersionless flat band whereas the other two dispersive bands are at Dirac points ${\bf k }= (\pm \frac{2\pi}{3a},0)$ as illustrated in Fig.\  \ref{FIG:1}. Within the first Brillouin zone, there are six of these Dirac points. The effective Hamiltonian near these points turns out to have  two dimensional Dirac points and reproduce the gapless linear band structure as that in graphene. The aim of this paper is to expand the Hamiltonian near these points, decouple the flat band and consider the effective Hamiltonian near the Dirac points and  investigate the effect due to irradiation. As in graphene, circularly polarized light irradiation opens a significant gap at these points and  therefore we will be able to calculate the exciton binding energy. However, in the presence of linearly polarized irradiation, the two gapless branches form the valence and conduction bands and do not separate to open up a gap, as it is the case in graphene. In order to obtain an effective Hamiltonian, we first  Taylor expand the Hamiltonian in Eq.\ (\ref{K_Hamil_F}), around ${\bf k_D} =( \frac{2\pi}{3a},0) $  by changing variables as ${\bf k } ={\bf k_D}+ {\bf q}$ assuming $|{\bf q}|<<|{\bf k_D}|$. The next step is to isolate the eigenvalue corresponding to the flat band. This is done by performing  a  unitary transformation of the resulting matrix. The  details  of this process were elegantly  presented in Ref. [\onlinecite{ciola}].

\medskip
\par
The convenient form of the resulting effective Hamiltonian near the Dirac point comes out to be

 \begin{equation}
 \hat{{\cal H}}( {\bf q})
   = \  \left(
 \begin{array}{cc}
  t_h +\hbar \upsilon_F q_1  & \hbar \upsilon_F q_2  \\
   \hbar \upsilon_F q_2 & t_h -\hbar \upsilon_F q_1   
      \end{array}
 \right) 
 \label{H_q}
\end{equation} 
which involves the following change of variables

\begin{eqnarray}
q_1 = \frac{1}{2}(q_x + \sqrt{3}q_y) \\ \nonumber  
q_2 = \frac{1}{2}( \sqrt{3}q_x -q_y) \   .
\end{eqnarray}
We emphasize that this is different from the Dirac Hamiltonian for graphene.  It can be shown in a straightforward way that the eigenvalues of the Hamiltonian in Eq.\ (\ref{H_q}) are given by  $\epsilon_\pm({\bf q})=t_h\pm \hbar \upsilon_F q $, exhibiting a linear dispersion near the Dirac point. 
In the forthcoming sections, we investigate the effects due to irradiation on the kagome energy bands near the Dirac points. For this, we employ the effective Hamiltonian in Eq.\  (\ref{H_q}).  

\medskip
\par

\section{Electron Dressed States}
 \label{sec3}

In general, the vector potential $\mbox{\boldmath$A$}^{(E)}(t)$ for  elliptical (clockwise) polarization can be written as

\begin{equation}
\label{ellipA}
\mbox{\boldmath$A$}^{(E)}(t) =
\left[  \begin{array}{c}
          A^{(E)}_x (t) \\
          A^{(E)}_y (t)
        \end{array}
\right] = \frac{E_0}{\omega} \left[
\begin{array}{c}
\cos \theta_p \cos (\omega t) - \beta \, \sin \theta_p \sin (\omega t) \\
\sin \theta_p \cos (\omega t) + \beta \,\cos \theta_p \sin (\omega t)
\end{array}
\right] \, 
\end{equation}
where $E_0$ is the electric field strength,   $\theta_p$ denotes the polarization angle of the optical field, measured with respect to the $x$ axis.

\medskip
\par
In the presence of an external electromagnetic field, the kagome Hamiltonian  can be decomposed into two parts: a wave vector-dependent part $\hat{{\cal H}}( \bf {q})$ in Eq.\ (\ref{H_q}) and a time-dependent part, $\hat{{\cal H}}(t)$. The time-dependent interaction Hamiltonian for circularly polarized light i.e., $(\beta =1)$ is given by

\begin{equation}
{\cal H}(t)
   = -c_0 \  \left(
 \begin{array}{cc}
\cos(\theta_p+ \omega t)& \sin(\theta_p+\omega t) \\
   \sin(\theta_p+\omega t) & -\cos(\theta_p+\omega t)  \\
 \end{array}
 \right) 
 \label{H_t}
\end{equation}
where we introduced the energy-dependent  variable $c_0\equiv  e \upsilon_F E_0/\omega$.  Now, the total Hamiltonian  $\hat{{\cal H}}^{tot}(t)  = \hat{{\cal H}}( {\bf q})+\hat{{\cal H}}(t)$ is time periodic and obeys the time-dependent Schr\"{o}dinger  equation given by

\begin{equation}
  \hat{{\cal H}}^{tot}(t) \Psi ({\bf q}, t) =
i\hbar\frac{\partial  \Psi ({\bf q}, t)}{\partial t} \ .
\label{SCHE}
\end{equation}
The solution to the Eq.\ \eqref{SCHE} is also periodic in time and given by  

\begin{equation} 
 \Psi ({\bf q}, t+T)=\text{e}^{-i {\cal H}_{eff}T/ \hbar} \Psi ({\bf q}, t)
 \label{PSI}
 \end{equation}
  where $\hat{{\cal H}}_{eff}$ is the Floquet effective Hamiltonian which is derived using the Floquet-Magnus expansion. In this expansion, the effective Hamiltonian can be expressed in the form

\begin{equation}
\hat{{\cal H}}_{eff} = \sum_{n=0} ^{\infty}  \hat{{\cal H}}^{(n)}_{eff}
\label{EFF}
\end{equation}
where $\hat{{\cal H}}^{(n)}_{eff} \sim \omega^{-n}$. 

\begin{equation}
\hat{{\cal H}}^{(0)}_{eff}= \frac{1}{T} \int_{0}^{T} dt\  \hat{{\cal H}}^{tot}(t)  
= \hat{{\cal H}}({\bf q})
\label{EFF1}
\end{equation}

\begin{equation}
\hat{{\cal H}} ^{(1)}_{eff}= \frac{1}{2 i \hbar T} \int_{0}^{T} dt_{1}\int_{0}^{t_{1}} dt_{2} \left[\hat{{\cal H}}^{tot}(t_{1}), \hat{
{\cal H}}^{tot}(t_{2})\right]
\label{EFF2}
\end{equation}

\medskip
\par 
\begin{eqnarray}
\hat{ {\cal H}}^{(2)}_{eff}&=& \frac{1}{3! (i \hbar )^2 T} \int_{0}^{T} dt_{1}\int_{0}^{t_{1}} dt_{2}\int_{0}^{t_{2}} dt_{3} \left(\left[\hat{
{\cal H}}^{tot}(t_{1}),\left[ \hat{
{\cal H}}^{tot}(t_{2}),{\cal H}^{tot}(t_{3})\right]\right] \right. \nonumber\\
& &\left. + \left[\hat{
{\cal H}}^{tot}(t_{3}),\left[ \hat{
{\cal H}}^{tot}(t_{2}), \hat{
{\cal H}}^{tot}(t_{1})\right]\right] \right)
\label{EFF3}
\end{eqnarray}
\medskip
\par 
The leading order correction to the Hamiltonian is the first order correction proportional to $1/\omega$,  given by Eq.\ \eqref{EFF2}.  This is calculated as

\begin{equation}
{\cal H}^{(1)}_{eff}
   =  \frac{ i c_0}{\hbar \omega} \left( 2 \hbar \upsilon_F q_1 \cos\theta_p + 2 \hbar \upsilon_F q_2 \sin\theta_p - c_0\right) \ \left(
 \begin{array}{cc}
 0 & 1 \\
 -1 & 0  
 \end{array}
 \right) \  ,
\label{Heff_1C}
\end{equation}
where, in this notation, $T=2\pi/\omega$. After a perturbation expansion, the total effective Hamiltonian is obtained and we have

\begin{equation}
\hat{{\cal H}}^{C}_{eff} = 
\left(
\begin{array}{cc}
t_h + \hbar \upsilon_F q_1 & 
\begin{array}{l}
\hbar \upsilon_F q_2 + \dfrac{i c_0}{\hbar \omega} \big( 2 \hbar \upsilon_F q_1 \cos\theta_p \\
\quad +\ 2 \hbar \upsilon_F q_2 \sin\theta_p - c_0 \big)
\end{array} 
\\[12pt]
\begin{array}{l}
\hbar \upsilon_F q_2 - \dfrac{i c_0}{\hbar \omega} \big( 2 \hbar \upsilon_F q_1 \cos\theta_p \\
\quad +\ 2 \hbar \upsilon_F q_2 \sin\theta_p - c_0 \big)
\end{array} 
& t_h - \hbar \upsilon_F q_1
\end{array}
\right)
\label{Heff_C}
\end{equation}
\medskip
\noindent
Solving the eigenvalue equation for this effective Hamiltonian, we obtain the quasiparticle energy spectrum for the kagome lattice under circularly polarized electromagnetic radiation   for arbitrary phase angle $\theta_p$. We have

\begin{eqnarray}
{\cal E}^{C}_{s}({\bf q}) &=& t_h + s\left\{ \hbar^2\upsilon^2_F q^2_1 +\hbar^2\upsilon^2_F q^2_2 +\left(\frac{  c_0}{\hbar \omega}\right)^2   \left( 2 \hbar \upsilon_F q_1 \cos\theta_p + 2 \hbar \upsilon_F q_2 \sin\theta_p - c_0\right)^2  \right\}^{1/2}    
\nonumber\\
&\approx&  t_h+s\frac{c_0^2}{\hbar\omega} -\hbar  s\upsilon_{Fx} q_x-\hbar s\upsilon_{Fy} q_y +s\frac{\hbar^2 (q_x^2+q_y^2)}{2m_{e,h}}
+\cdots  
\label{E_eff_C}
\end{eqnarray}
with

\begin{equation}
\upsilon_{Fx}\equiv \frac{c_0\upsilon_F}{\hbar\omega}\left[\cos\theta_p+\sqrt{3}\sin\theta_p\right]\  ,
 \ \ \  \ \upsilon_{Fy}\equiv    \frac{c_0\upsilon _F}{\hbar \omega}\left[\sqrt{3}\cos\theta_p -\sin\theta_p\right]\  ,  m_{e,h}=  \frac{c_0^2}{\upsilon_F^2\hbar\omega}            .
\label{E_eff_C2}
\end{equation}
It is remarkable that the group velocity is anisotropic but the effective mass is not.     Clearly,    the application of circularly polarized irradiation opens up an energy gap at the Dirac point. This gap is given by

\begin{equation}
{\cal E}^{C}_{g} =  \frac{2  c^2_0}{\hbar \omega}  
\label{E_C_gap}
\end{equation} 
where $ {\cal E}^{C}_{+}({\bf  q}) -{\cal E}^{C}_{-}({\bf q }) = {\cal E}^{C}_{g}+ O(q)$.

\medskip
\par

The corresponding eigenvectors for the effective Hamiltonian can be expressed in terms of  the energy gap as 

\begin{equation}
\label{wavevector_C}
\Psi_{C; s} =  \left(
\begin{array}{c}
\frac{\hbar \upsilon_F q_1 + s {\cal E}^{C}_{g}/2}{\text{D}} \\ \\
1          
\end{array}
\right) \  .
\end{equation}
where

\begin{equation}
\text{D} = \hbar \upsilon_F q_2 - \frac{i c_0} {\hbar \omega} 2 \hbar \upsilon_F (q_1 \cos\theta_p + q_2 \sin\theta_p) - \frac{i c^2_0}{\hbar \omega}. 
\end{equation}
\medskip
\par

\begin{figure} 
\centering
\includegraphics[width=0.48\textwidth]{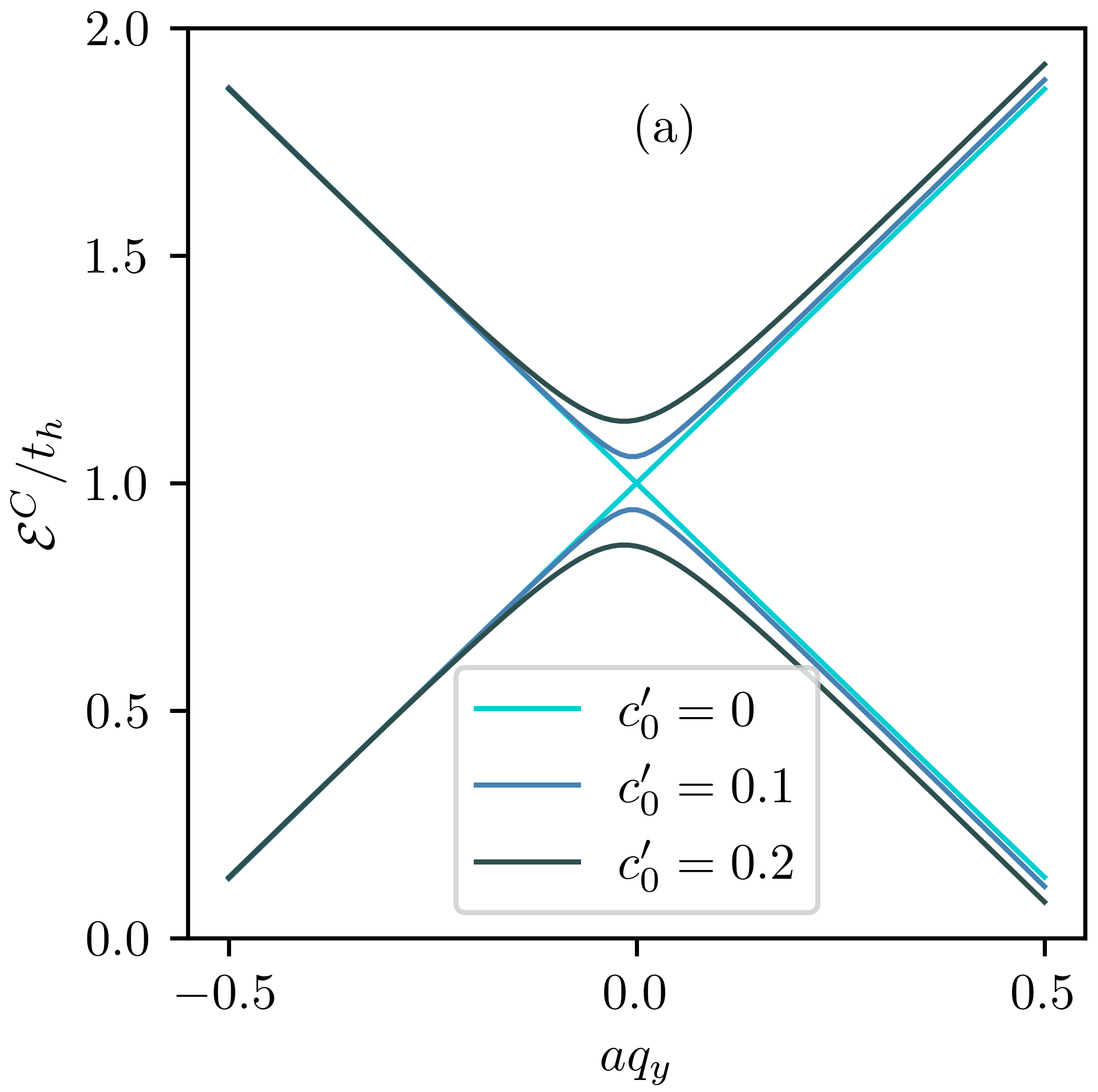}
\includegraphics[width=0.48\textwidth]{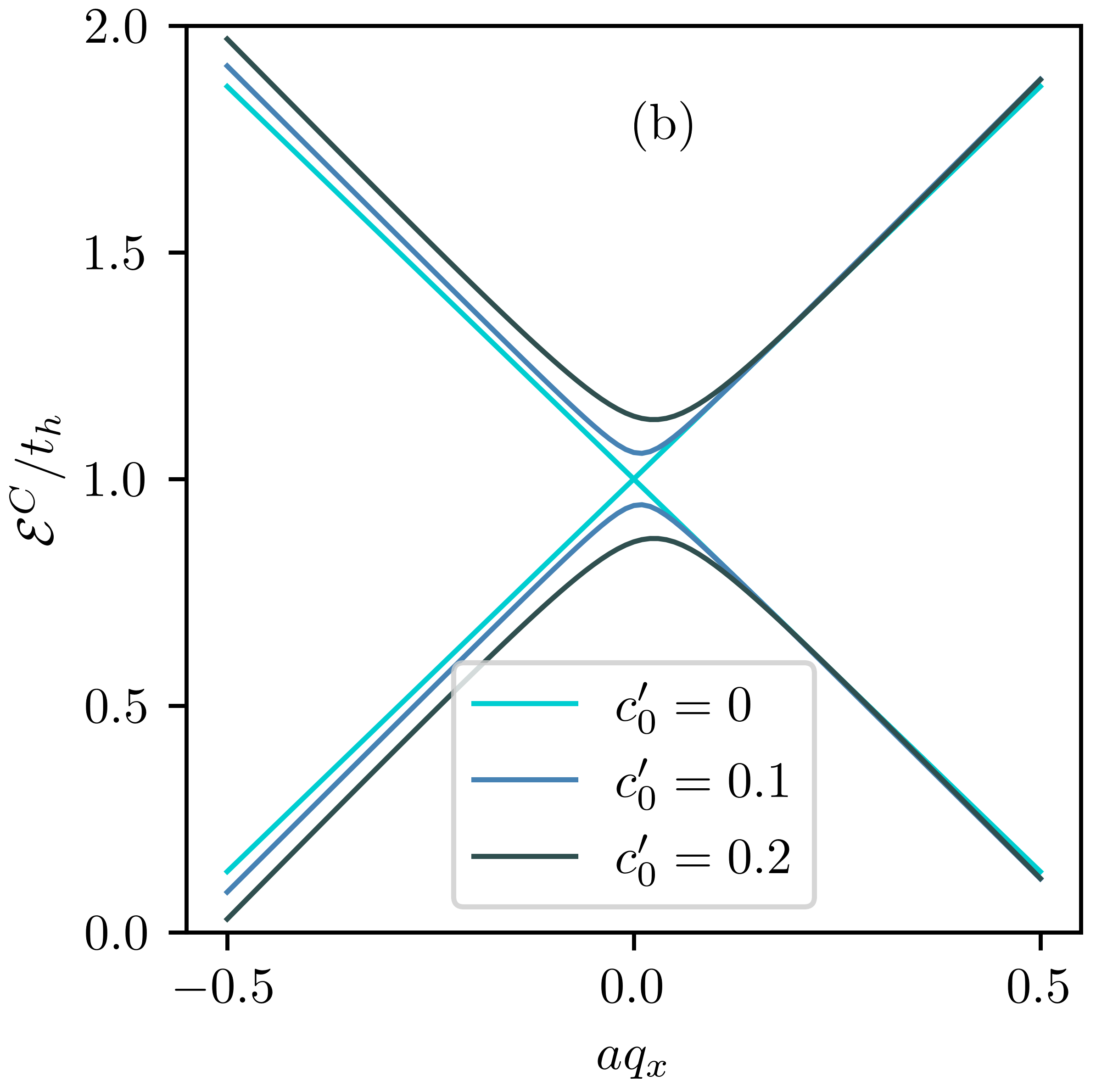}
\caption{(Color online) 
Energy dispersion of the kagome lattice as a function of $aq_y$ for $q_x =0$ in (a) and as a function of $aq_x$ for $q_y =0$ in (b). Here,  $a$ is the lattice spacing or the bond length between atoms. The three curves corresponding to $c^{\prime}_0 = c_0/\hbar\omega$ represent the dispersion line in the absence of irradiation ($c^{\prime}_0 =0$) and in the presence of circularly polarized light irradiation for  two chosen values of irradiation parameter $c^{\prime}_0 =0.1$ and $c^{\prime}_0 =0.2$, respectively. The energy is plotted in unit  of $t_h$. These results are for high frequency photons, i.e, $\hbar\omega \gg  c_0$ and $\hbar\omega \gg \hbar\upsilon_F /a$. The energy bands are anisotropic in two perpendicular momentum directions and open a significant gap in the presence of irradiation.} 
\label{FIG:2}
\end{figure}

\medskip
\par

\begin{figure} 
\includegraphics[width=0.6\textwidth]{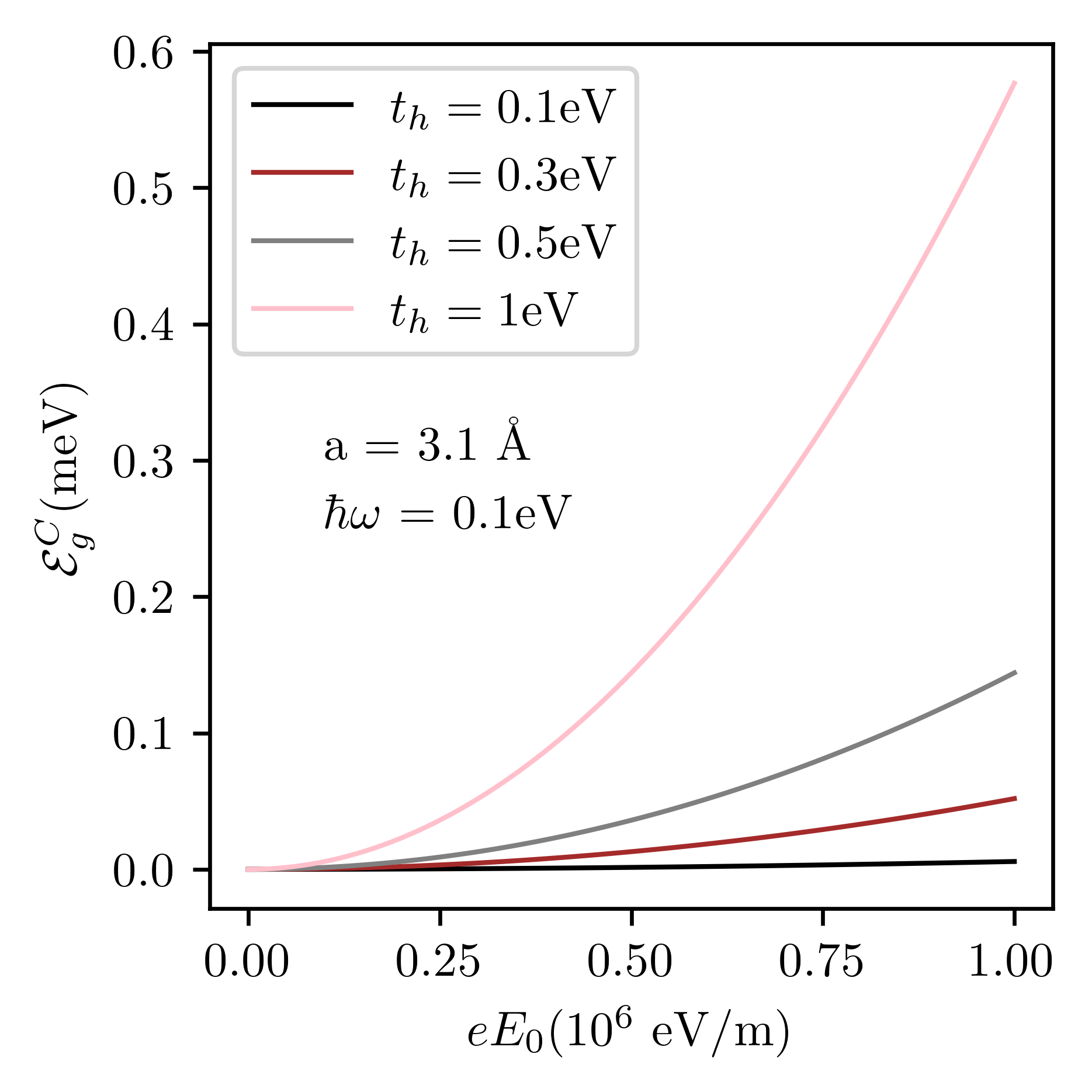}
\caption{(Color online) 
Energy band gap between two dispersive bands  in meV as a function of the electromagnetic field intensity for the external irradiation with circular polarization.     Four different value of hoping constant $t_h$ are chosen for reference.  These results are for photon  frequency corresponding to 0.1 eV and lattice spacing $a=3.1\AA$, as shown in the figure. The band gap, which is tunable by external irradiation, is a crucial  quantity for studying the excitonic properties of the material. Study of excitons and superfluidity in kagome lattice  is the main objective of this work.}
\label{FIG:3}
\end{figure}

\begin{figure} 
\includegraphics[width=0.6\textwidth]{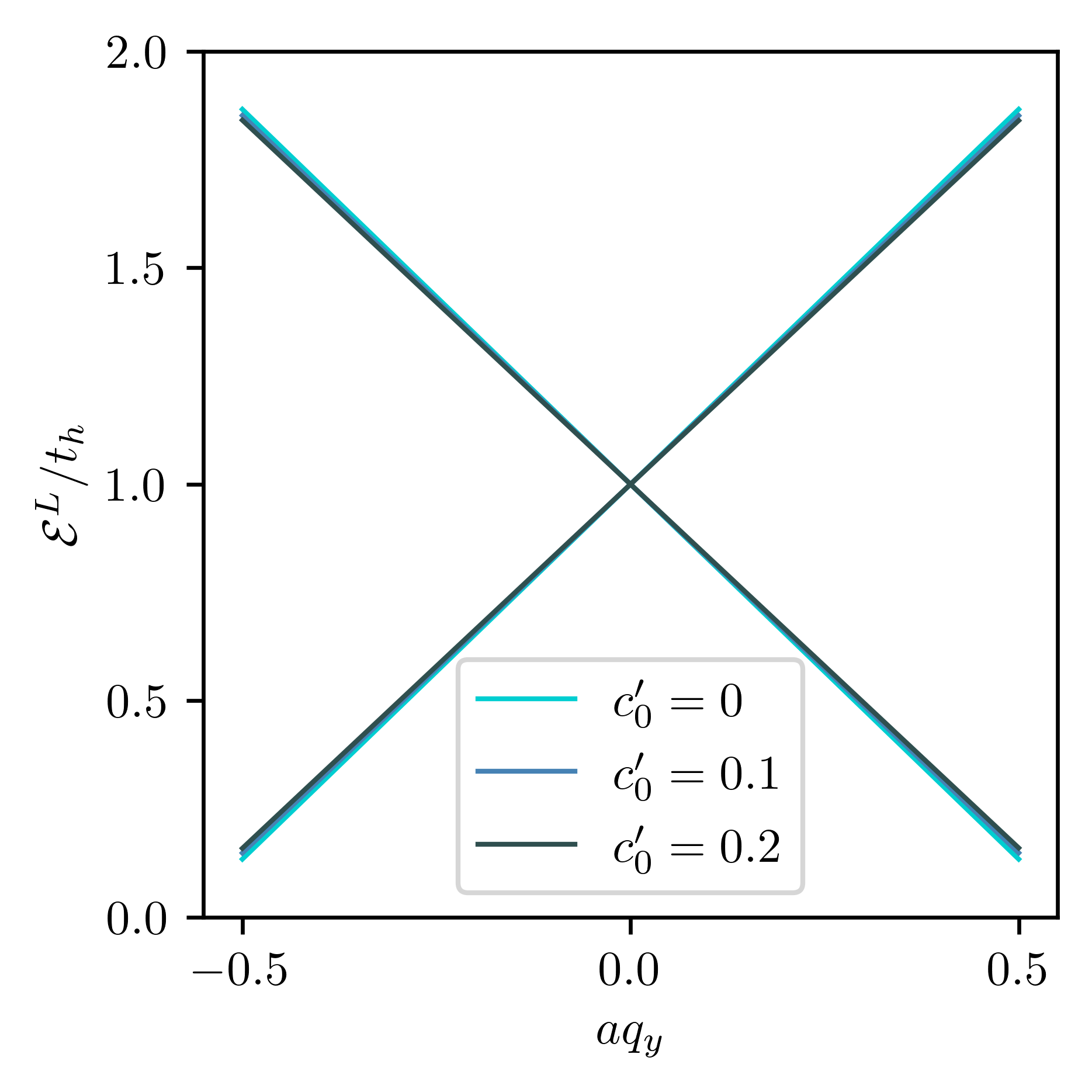}
\caption{(Color online) 
Energy dispersion for the kagome lattice as a function of wave vector $aq_y$ for $q_x =0$,  in terms of the lattice spacing $a$.  The three curves corresponding to $c^{\prime}_0$ = 0, 0.1 and 0.2  fall on top of each other. Therefore, these bands  are not modified  in the presence of ``linearly polarized” irradiation and are isotropic in two perpendicular momentum directions. The energy is plotted in units of $t_h$. These results are for high frequency photons, i.e, $\hbar\omega \gg c_0$ and $\hbar\omega \gg \hbar\upsilon_F /a$. } 
\label{FIG:4}
\end{figure}

\medskip
\par

In Fig.~\ref{FIG:2}, we compare the energy dispersions of kagome materials in the presence and absence of circularly polarized irradiation in two different momentum directions. The anisotropy and gap opening are clearly displayed.  From Eq.\ (\ref{E_C_gap}), it is clear that the gap depends on the external field intensity $E_0$, frequency of irradiation $\omega$ and the material parameters such as $a$ and $t_h$ through the relation $\hbar \upsilon_F = \sqrt{3} a t_h$. In Fig.~\ref{FIG:3}  the variation of the band gap as a function of the field intensity for chosen photon frequency and lattice parameter $a$ is presented. Four arbitrary values of hopping constant $t_h$ are selected to represent different kagome materials, for example. For photons of high frequency about 100 THZ and chosen values of $a$ and $t_h$ for kagome lattice, a band gap of few meV can be opened with the external electromagnetic field of intensity $\sim 10^6$ V/m. In  later sections,  we explore the excitonic properties of this tunable gapped kagome lattice.  
\medskip
\par

Now, for linearly polarized light for $\beta = 0$, the first-order correction to the Hamiltonian $\hat{H}^{(1)}_{eff}$ becomes zero. Therefore, the  leading order correction to the Hamiltonian is second order proportional to $1/\omega^2$, which has been calculated to be

\begin{eqnarray}
\hat{{\cal H}}^{(2)}_{eff}
   &=&\frac{1}{\hbar^2 \omega^2} \left\{  \frac{10 c_0 \hbar^2 \upsilon^2_F}{3}\ \left(
 \begin{array}{cc}
 q^2_2 \cos\theta_p - q_1 q_2 \sin\theta_p &  q^2_1 \sin\theta_p - q_1 q_2 \cos\theta_p  \\
 q^2_1 \sin\theta_p - q_1 q_2 \cos\theta_p& -q^2_2 \cos\theta_p + q_1 q_2 \sin\theta_p
 \end{array}
 \right) \right.
 \nonumber \\
 & & \left. -  c^2_0\hbar\upsilon_F \left(
 \begin{array}{cc}
 q_1 \sin^2\theta_p- q_2 \sin\theta_p \cos\theta_p & q_2 \cos^2\theta_p- q_1 \sin\theta_p \cos\theta_p \\
q_2 \cos^2\theta_p- q_1 \sin\theta_p \cos\theta_p &  -q_1 \sin^2\theta_p+q_2 \sin\theta_p \cos\theta_p
 \end{array}
 \right) \right\} .
\label{H_eff_L2}
\end{eqnarray}
  
With this correction to the unperturbed Hamiltonian $\hat{{\cal H}} ({\bf q}) $,  the Floquet effective Hamiltonian for linearly polarized light can be written in simple matrix form as

\begin{equation}
\hat{{\cal H}}^{(L)}_{eff}   = \left(
 \begin{array}{cc}
 t_h + L_1({\bf  q})&  L_2({\bf  q}) \\
 L_2({\bf  q})&  t_h - L_1({\bf  q})
 \end{array}
 \right)
\label{H_eff_L}
\end{equation}
where 
 
\begin{equation}
L_1({\bf  q}) = \hbar\upsilon_F q_1 + \left(\frac{10 c_0 \hbar^2 \upsilon^2_F}{3\hbar^2 \omega^2}\right) (q^2_2 \cos\theta_p - q_1 q_2 \sin\theta_p ) - \frac{c^2_0 \hbar\upsilon_F}{\hbar^2 \omega^2}(q_1 \sin^2\theta_p- q_2 \sin\theta_p \cos\theta_p)
\label{L_1}
\end{equation} and 
\begin{equation}
L_2 ({\bf  q})= \hbar\upsilon_F q_2 + \left(\frac{10 c_0 \hbar^2 \upsilon^2_F}{3\hbar^2 \omega^2}\right) ( q^2_1 \sin\theta_p - q_1 q_2 \cos\theta_p) - \frac{c^2_0 \hbar\upsilon_F}{\hbar^2 \omega^2}(q_2 \cos^2\theta_p- q_1 \sin\theta_p \cos\theta_p )
\label{L_2} .
\end{equation}
Solving the eigenvalue equation for this effective Hamiltonian, we obtain the quasiparticle energy spectrum for the kagome lattice under linearly polarized electromagnetic radiation as

\begin{equation}
{\cal E}^{L}_{s} ({\bf  q}) = t_h + s\sqrt{ L^2_1({\bf  q}) + L^2_2({\bf  q})  }     
\label{E_eff_L}.
\end{equation}
The corresponding  wave vectors  are
\begin{equation}
\label{wavevector_L}
\Psi_{L; s} =  \left(
\begin{array}{c}
\frac{L_1 + s\sqrt{ L^2_1+ L^2_2 } }{L2} \\ \\
1          
\end{array}
\right) 
\end{equation}

\medskip
\par
\noindent
Clearly, since both $L_1({\bf  q})$ and $L_2({\bf  q})$ vanish when ${\bf  q}=0$, there is no band gap opening at the Dirac point when linearly polarized light is applied to the kagome lattice. Unlike the case of circular polarization, a gap does not open near the Dirac points and the dispersion of the bands also do not change. The dispersion for linearly polarized irradiation is shown in Fig. \ref{FIG:4}.

\section{Interacting Electron-hole Pair}
 \label{sec4}

In this section, we aim to calculate the exciton binding energy of the kagome lattice in the presence of circularly polarized light irradiation. The exciton is a bound state of an electron-hole pair whose effective mass and group velocity are presented above in Eq.\ (\ref{E_eff_C2}).  Excitons in different materials have unique properties. They do not present themselves in all kinds of materials but only in some semiconductors and TMDCs. If we want to understand the optical properties of those materials, we should be aware of their excitonic spectra. The exciton binding energy is the energy required to decompose an exciton into its constituent electron and hole charge carriers.   

\medskip
\par
The electron-hole pair is bound together by the Coulomb interaction,  thereby resulting in a two dimensional hydrogen-like atom. However, the two-body Hamiltonian consists of a Schr\"{o}dinger part  with a massive particle and a Dirac part whose dependence on momentum is linear.  In the case of 2D materials, we can calculate the energy spectrum of the exciton by solving the 2D  Schr\"{o}dinger-Dirac-like Hamiltonian equation.

\medskip
\par

\begin{figure} 
\centering
\includegraphics[width=0.6\textwidth]{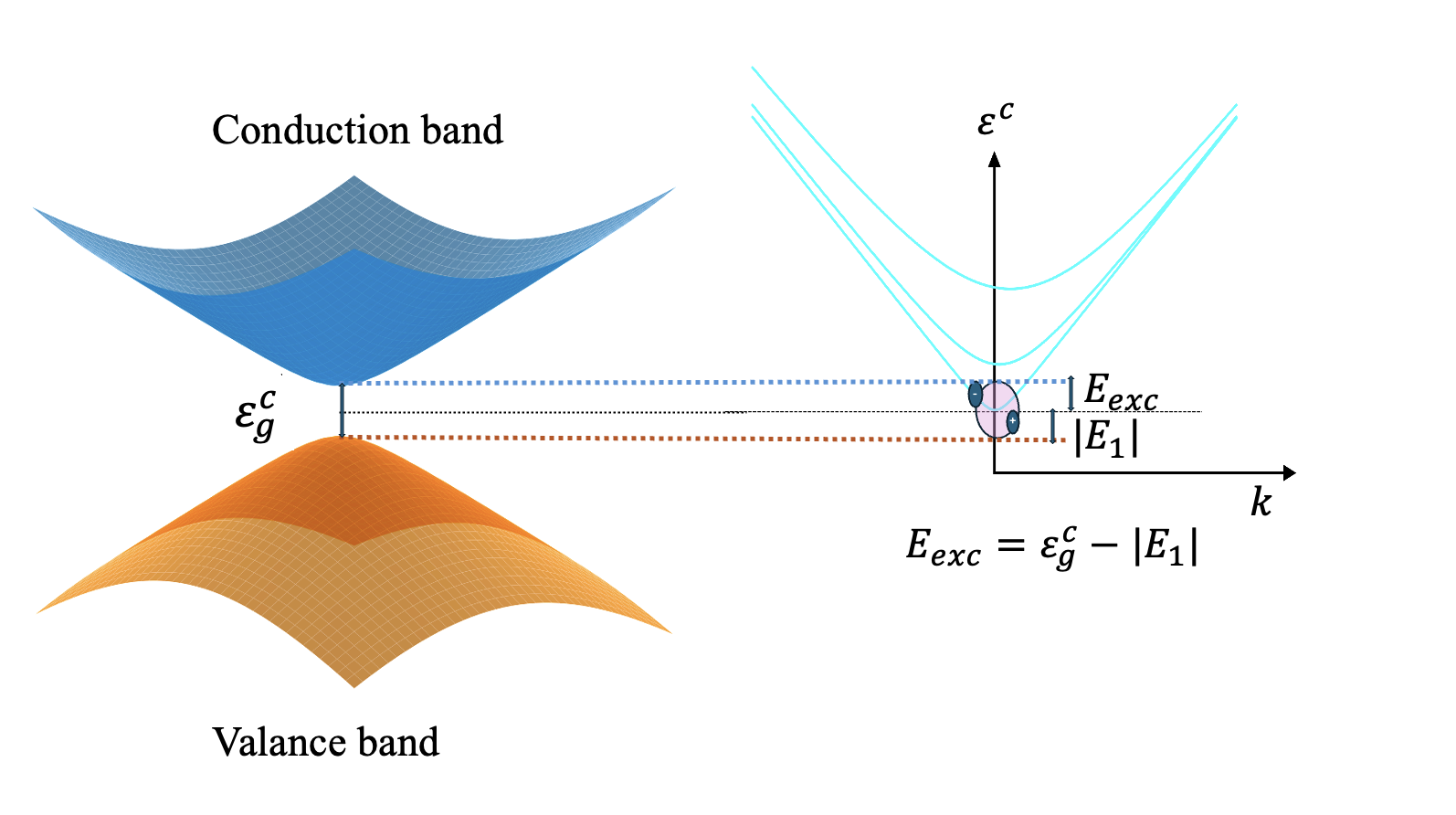}
\caption{(Color online) 
Schematic representation of gapped kagome material in the presence of circularly polarized irradiation. The exciton energy is defined as the difference between the energy band gap and the absolute value of the ground state energy of the exciton spectrum. }
\label{FIG:5}
\end{figure}

 Based on this model, the exciton binding energy is  defined as the absolute value of the exciton ground state energy, $|E_1|$. The exciton energy is the  difference between the energy gap and  the absolute value of the exciton ground state energy. That is,  
 $E_{exc} = {\cal E}^{C}_{g} - |E_1|$  
depicted schematically in Fig.\  \ref{FIG:5}. 
The exciton ground state energy, $E_1$,  is determined    by solving the following equation 
in 2D for an electron and a hole in monolayer kagome lattice  

\medskip
\par

The electron (charge $-e$) and hole (charge $+e$) with effective  mass $m_{e,h}$ are separated by distance $\rho$.  They  are attracted  by the Coulomb potential $V(\rho) = -e^2/\varepsilon_s \rho$ where $\varepsilon_s\equiv  4\pi\epsilon_0\epsilon_b $ with $\epsilon_b $ denoting the background dielectric constant. 
The motion of the interacting electron-hole pair is   governed by

 \begin{eqnarray}
 &&\left[-\frac{\hbar^2}{2m_{e,h}} \left(  \frac{\partial^2}{\partial x^2_1}  +  \frac{\partial^2}{\partial y^2_1}\right)    -\frac{\hbar^2}{2m_{e,h}} \left(  \frac{\partial^2}{\partial x^2_2}  +  \frac{\partial^2}{\partial y^2_2}\right) \right. 
\nonumber\\
&-i\hbar& \left. \left( \upsilon_{Fx}\frac{\partial}{\partial x_1}  + \upsilon_{Fy}\frac{\partial}{\partial y_1}\right)
-i\hbar  \left( \upsilon_{Fx}\frac{\partial}{\partial x_2}  + \upsilon_{Fy}\frac{\partial}{\partial y_2}\right)
\ + V(\rho) \right] \psi({\bf r}_1,{\bf r}_2) 
\nonumber\\
& =&  {\cal E}
\psi({\bf r}_1,{\bf r}_2)
 \label{eq1}
\end{eqnarray}

\medskip
\par

Noting that the potential energy is a function only on the relative coordinates, i.e., $V = V(x_1 - x_2, y_1 - y_2)$, let us define the relative coordinates $x, y$ and coordinates of the center of mass $X,Y$ by $ x= x_1-x_2, 
 y= y_1- y_2 $ and $MX = m_{e,h} (x_1 +  x_2),  MY = m_{e,h} (y_1 +  y_2)$ with $M = 2m_{e,h}$ being the total mass of the system. With this coordinate transformation, the Schr\"{o}dinger-Dirac equation now becomes,

\begin{eqnarray}
&& \left[-\frac{\hbar^2}{2M} \left(  \frac{\partial^2}{\partial X^2}  +  \frac{\partial^2}{\partial Y^2}\right)    -\frac{\hbar^2}{2\mu} \left(  \frac{\partial^2}{\partial x^2}  +  \frac{\partial^2}{\partial y^2}\right)  
\right. 
\nonumber\\
&-&
\left. i\hbar \frac{3}{2} \left(\upsilon_{Fx}\frac{\partial}{\partial X}+
 \upsilon_{Fy}\frac{\partial}{\partial Y} 
\right) 
+ V(x,y) \right]\psi ({\bf r}_1,{\bf r}_2)=    {\cal  E}\psi({\bf r}_1,{\bf r}_2) 
 \label{eq2}
 \end{eqnarray}
 where $\mu = \frac{m_{e,h} }{ 2}$ is the reduced mass of the electron-hole pair.

\medskip
\par

Proceeding with a separation of variables, we have     , relative motion of interacting particle with reduced mass $\mu$ and free particle motion of particles with centre of mass $M$.  That is, we can write $\psi(x, y, X, Y) = u(x,y) U(X,Y)$.  This results into separate differential equations

 \begin{equation}
 -\frac{\hbar^2}{2\mu} \nabla^2 u + V u = E u
 \label{eq3}
 \end{equation}
 and

\begin{eqnarray}
&& 
\left[-\frac{\hbar^2}{2M} \left(  \frac{\partial^2}{\partial X^2}  +  \frac{\partial^2}{\partial Y^2}\right)
\right. 
\nonumber\\
&-&
\left. \frac{3}{2} i\hbar \left(\upsilon_{Fx}\frac{\partial}{\partial X}+  \upsilon_{Fy}\frac{\partial}{\partial Y} 
\right) 
     \right]   U(X,Y)=  
E^{\prime} U(X,Y) \    .
\label{eq4a}
 \end{eqnarray}
We are interested in the relative motion of a pair of particles which have  reduced mass $\mu$ in an external potential $V(\rho)$. One can find  the  solution of Eq.\ (\ref{eq3})  as \cite{AJP}

\begin{equation}
E_n = - \frac{\mu (e^2/\varepsilon_s)^2}{2\hbar^2 (n-1/2)^2}
\label{eq14}
\end{equation}
for $n= 1,2,3,\cdots$ .

\medskip
\par

Therefore, from Eq.\ (\ref{eq14}) and  substituting  the quasiparticle reduced mass 
$\mu =    \frac{ c^2_0}{2 \ \upsilon^2_F\hbar \omega}$,
the exciton binding energy of  kagome lattice is calculated as $|E_{n=1}| =  \frac{ c^2_0 (e^2/\varepsilon_s)^2}{4\hbar^3 \upsilon^2_F\omega/4  }$.   
Accordingly, the exciton energy for the kagome lattice is given by

\begin{figure} 
\centering
\includegraphics[width=0.45\textwidth]{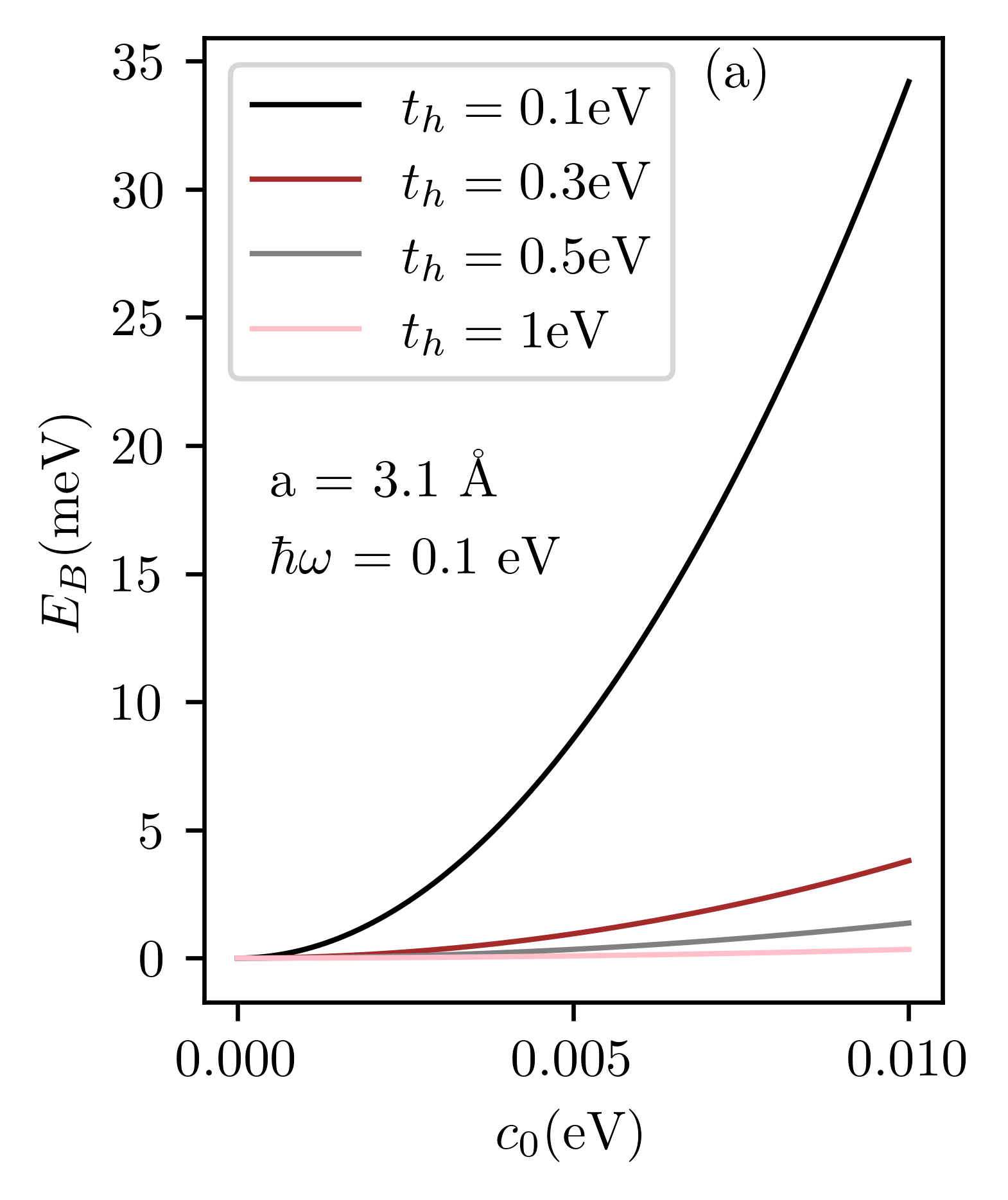}
\includegraphics[width=0.45\textwidth]{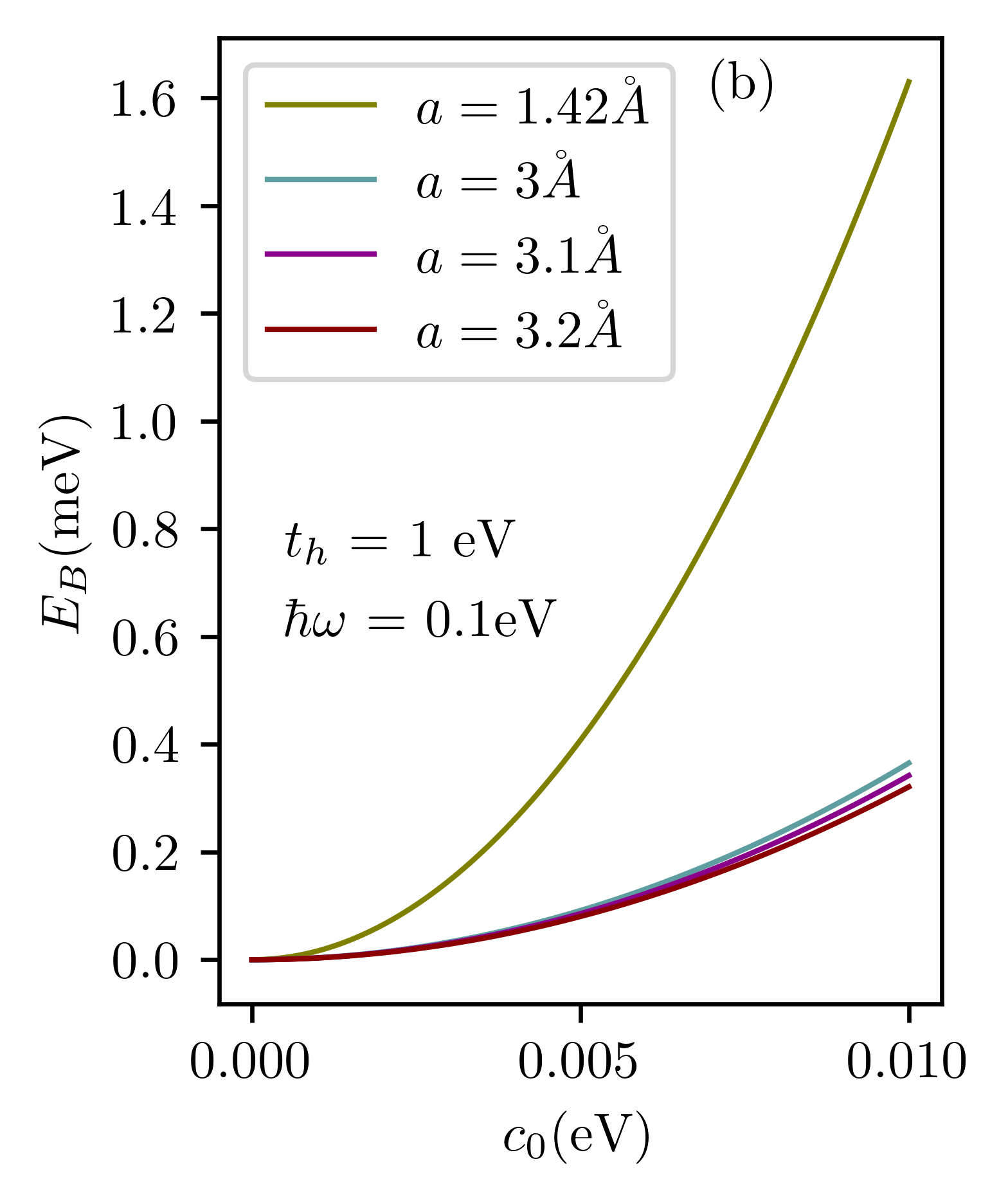}
\caption{(Color online) 
Exciton binding energy of monolayer  kagome lattice, defined as $E_{B} = | E_{1}| $   in terms of the ground state excitonic energy,  as a function of the irradiation intensity $c_0$. We apply an electromagnetic field of frequency 100 THz  and  four chosen values of the hopping parameter $t_h$ in  (a) and four chosen values of lattice constant $a$   with $t_h$ = 1 eV in  (b).The binding energy is expressed in unit of meV and all other energies are in unit of  eV as indicated. }
\label{FIG:6}
\end{figure}

\begin{equation}
E_{exc} = \hbar\omega \left( \frac{c_0}{\hbar\omega} \right)^2 \left[2  - \frac{e^4} {\hbar^2 \upsilon^2_F \varepsilon_s^2}   \right]\ .
\label{eq15}
\end{equation}
\noindent
This expression gives the exciton energy which depends on the photon frequency $\omega$ and the irradiation intensity $c_0$. We note that for the  kagome lattice, the Fermi velocity  $\upsilon_F $ is related to the  hopping parameter $t_h$ through  $\hbar \upsilon_F = \sqrt{3} a t_h$. The exciton energy in Eq.\ (\ref{eq15}) turns negative when the binding energy exceeds the band gap, i.e.,  \(\hbar\upsilon_F< \frac{e^2}{\sqrt{2}\varepsilon_s}\), which may also be expressed in terms of the hopping energy as \( t_h< \frac{e^2}{\sqrt{6}\varepsilon_s\ a}\) which gives the condition for a phase transition in a kagome lattice from the semiconducting phase to the excitonic insulating (EI) phase.  If the exciton binding energy is less than the band gap, the ground state of the system is the semiconductor phase, whereas if the exciton binding energy is greater than  the band gap, the ground state of the system is the EI phase.  The dielectric constant $\epsilon_b$ for 2D materials varies between 2.0 and 10.  Our numerical results for the exciton binding energy for monolayer kagome lattice with arbitrarily chosen values of $a$ and $t_h$ as a function of irradiation intensity  are presented in Fig.\  \ref{FIG:6}.  The background dielectric constant was chosen to be that of  $h-$BN,  i.e., $\epsilon_b = 4.58$.  The quadratic dependence of the exciton binding energy on the irradiation intensity is clearly shown in the plots. For the chosen parameters in Fig.\  \ref{FIG:6}(a), the binding energy is greater than energy gap for $t_h = 0.1$eV and 0.3 eV, therefore the ground state of the system for these values of $t_h$ is in the EI phase. However for $t_h = 0.5$ eV and 1.0 eV, the binding energy is less than the energy gap and hence the ground state of the system is in the semiconducting phase. Similarly, in Fig.\  \ref{FIG:6}(b), for all the chosen values of $a$,  $t_h = $1eV$> \frac{e^2}{\sqrt{6}\varepsilon_s\ a}$ and hence the ground state of the system is in superconducting phase.

 \section{Dipolar excitons in double layer  kagome lattice}
 \label{sec5}
 
    Different types of excitons may exist in semiconducting materials. They include dark excitons, bright excitons, and dipolar excitons \cite{dipolar4}.   Dark excitons are those which do not absorb or emit a photon directly. They require phonon assisted transition for momentum conserve while creating and annealing them. However, bright excitons can be created by absorbing photons and can be destroyed by releasing them directly. Dark excitons have been observed in tungsten (W)-based  monolayer TMDCs while bright excitons have been observed in molybdenum (M)-based monolayer TMDCs.\cite{arora_1, arora_2}. Dipolar excitons exist when electrons in one layer and holes in another layer are bound together by an electrostatic potential, resulting in permanent electric dipole moments. In this section, we investigate the energy spectrum of such dipolar excitons in double layer kagome lattice. 
    
\medskip
\par
Similar to Eq. \  (\ref{eq3}),  the Schr\"{o}dinger-Dirac like equation for relative motion of an electron-hole pair located in two parallel planes separated by distance $D$ and bound by an attractive Coulomb potential $V(r_{\parallel},D)=-e^2/ \epsilon_s \sqrt{r^2_{\parallel} + D^2} $, in Cartesian coordinates, can be written as,

\begin{equation}
\left[-\frac{\hbar^2}{2\mu}\left(  \frac{\partial^2}{\partial x^2}  +  \frac{\partial^2}{\partial y^2}\right)  + V(x,y,D) \right]  \varphi(x,y)= E^{\prime\prime} \varphi(x,y).
\label{eq16}
\end{equation}  
Assuming that $r_{\parallel} \ll D$, we expand the potential $V(r_{\parallel},D)$ as a Taylor series and retain only the first two terms in the expansion. We have

\begin{equation}
V(r_{\parallel}, D) \approx -V_0 + \beta r^2_{\parallel}  \  ,
\label{eq17}
\end{equation}
where

\begin{equation}
 r^2_{\parallel} = x^2 +y^2,  \, \, \, \, \, \, V_0 \equiv \frac{e^2}{  \epsilon_s D},  \,\,\,\,\,\,\,\,\,\,  \beta \equiv \frac{e^2}{ 2\epsilon_s D^3} \ .
\label{eq18}
\end{equation}
With the use of Eqs.\ (\ref{eq17}) and (\ref{eq18}) in Eq.\ (\ref{eq16}),  and with the separation of variables $\varphi(x,y) =\varphi(x)\varphi(y) $, one can obtain two independent equations for one dimensional simple harmonic oscillator along the $x$ and $y$ directions as follows:

\begin{equation}
-\frac{\hbar^2}{2\mu} \frac{d^2 \varphi(x)}{dx^2}    + \beta x^2 \varphi(x) =\left( E^{\prime\prime}_{x} + \frac{V_0}{2}\right) \varphi(x)
\label{eq19}
\end{equation}  
and 

\begin{equation}
-\frac{\hbar^2}{2\mu} \frac{d^2 \varphi(y)}{dy^2}    + \beta y^2 \varphi(y) =\left( E^{\prime\prime}_{y} + \frac{V_0}{2}\right) \varphi(y)
\label{eq20}
\end{equation}  
such that $E^{\prime\prime} = E^{\prime\prime}_{x}+ E^{\prime\prime}_{y}$.

The solutions of these equations can be found from one dimensional harmonic oscillator equations, which are 

\begin{equation}
E^{\prime\prime}_{n,x}  = E^{\prime\prime}_{n, y}  = -\frac{V_0}{2}+ \left(n+\frac{1}{2}\right)\hbar \sqrt{\frac{2\beta}{\mu}} 
\label{eq21}
\end{equation}  
 and 

\begin{equation}
\varphi(x) = \left(\frac{\alpha}{\pi}\right)^{1/4} \frac{1}{\sqrt{2^{n} n!}} e^{-\alpha x^2 /2} H_n(\sqrt{\alpha} \, x) ,\,\,\,\,\,\,\,\,\,\,\,\,\,\,
\varphi(y) = \left(\frac{\alpha}{\pi}\right)^{1/4} \frac{1}{\sqrt{2^{n} n!}} e^{-\alpha y^2 /2} H_n(\sqrt{\alpha} \, y)
\label{eq22}
\end{equation}
where $\alpha = \sqrt{2\beta\mu}/\hbar$,  $H_n(\sqrt{\alpha} \, x/y) $ is a Hermite polynomial of order $n$ and $n= 0, 1,2,3,\cdots$ is a quantum number. This gives the energy spectrum of dipolar exciton  \(E^{\prime\prime}_{n} = -V_0 + 2\left(n+\frac{1}{2}\right)\hbar \sqrt{\frac{2\beta}{\mu}}\). 
\medskip
\par
Now the exciton energy corresponding to this energy spectrum is given by

\begin{equation}
\boxed{E^{\prime\prime}_{exc} = {\cal E}^{C}_{g} -  E^{\prime\prime}_0 = \frac{2c^2_0}{\hbar\omega} -\left| - V_0 + \hbar \sqrt{\frac{2\beta}{\mu} } \right|}\ ,
\label{eq23}
\end{equation}
with the values for $V_0$ and $\beta$ given in Eq.\ (\ref{eq18}).
For the exciton binding energy, we have explicitly

\begin{equation}
|E_0|= \left|-\frac{e^2}{\varepsilon_s  D}  +t_h\left(\frac{a}{D}
\right)\left\{6\left(\frac{e^2}{\varepsilon_s  D} \right)
\left(\frac{\hbar\omega}{c_0^2}\right)  
   \right\}^{1/2}
\right|\  .
\label{ground2}
\end{equation}

\medskip
\par

This calculation shows that the dipolar exciton binding energy  depends inversely  on $c_0$ and also depends on specific material parameters such as $a$ and $t_h$ for chosen interlayer separation.  However, the variation of energy with separation $D$ for fixed intensity of the irradiation is non-monotonic. It depends on a  balance between the algebraic terms. We have $|E_0(D)| $  decreasing, reaches a minimum when  $D = (\hbar^2\upsilon^2_F / c^2_0) (2\hbar\omega \epsilon_s / e^2)$, then increasing slightly before eventually decreasing again toward zero. In general,  $|E_0(D)|\to \infty $ as $D \to 0$ and $ |E_0(D)|\to 0 $ as  $D\to \infty$.  The condition for large interlayer separations is that $D$  should be $\gg$  $\rho_{0}(D)$.  Here, $\rho_{0} (D) =\left(8r_{0}\right)^{1/4}D^{3/4}$ is the radius of a dipolar exciton along the plane of the layers, where $r_{0} = \hbar^{2}\varepsilon_s/\left(4\mu  e^{2}\right)$ is the Bohr radius of the 2D direct exciton in a monolayer, $\mu = m_{e,h}/2$ is the reduced  exciton mass, and $m_{e,h} = c_{0}^{2}/\hbar \omega \upsilon_{F}^{2}$ is the mass of an electron and a hole~\cite{Nishanov}.  The condition mentioned above for large $D$ is as follows: $D \gg 8 r_{0}$, which implies that

\begin{equation}
c_{0}^{2} \gg   4\left(\frac{\hbar\omega}{e^2/\varepsilon_s}  \right) \frac{(\hbar\upsilon_F)^2}{D} 
 \  .
\label{ineq1} 
\end{equation}

\begin{figure} 
\centering
\includegraphics[width=0.45\textwidth]{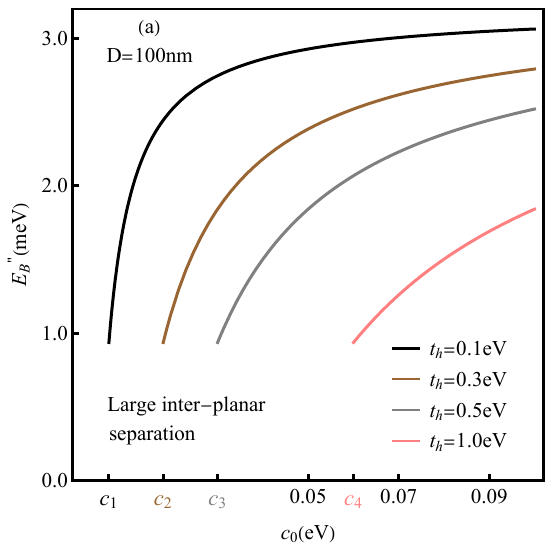}
\includegraphics[width=0.45\textwidth]{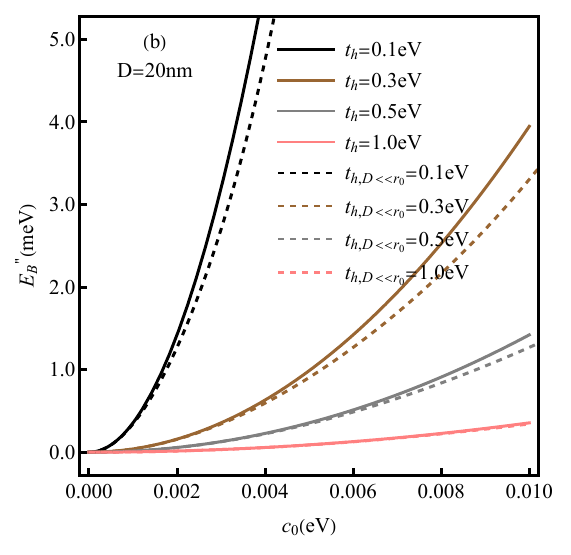}
\caption{(Color online) 
The binding energy as a function of the irradiation intensity parameter $c_0$ for a pair of kagome lattices with large separation $D=100$ nm $\gg 8r_0$ in (a) and  for small interplanar separation $D=20$ nm $\ll r_0$ in (b).   The plots are for frequency of the EM field chosen as $\omega=100$ THz,  and  four values of the hopping parameter $t_h$.  The lower bounds for the curves in (a) are denoted by $c_1,c_2, c_3$ and $c_4$ where $r_0 = D$. Binding energy is in unit of meV and all other energies are in unit of eV as indicated. We compare the exciton binding energy of monolayer (full curves) and double layer (dotted curves) in (b) for  $D=$ 20 nm with $D\ll  r_0$.} 
\label{FIG:7}
\end{figure}

\medskip
\par
For the case of small $D$, we imply that $D \ll r_{0}$. We can expand the electron-hole interaction potential $V(r)$ as a Taylor series with respect to the small parameter $D \ll r_{0}$ in the following way:
\begin{equation}
V(r) = - \frac{ e^{2}}{\varepsilon_s \sqrt{r_\parallel^{2} +
D^{2}}} \approx - \frac{ e^{2}}{\varepsilon_s  r_\parallel} + \frac{
e^{2}D^{2}}{2 \varepsilon_s r_\parallel^{3}}\   .
\label{Vrexp} 
\end{equation}

We will now obtain the exciton binding energy by applying perturbation theory:
\begin{equation}
E^{\prime\prime}_{B} =  E_{B}  - \left\langle \varphi (r_\parallel)
\left| \frac{ e^{2}D^{2}}{2 \varepsilon_s r_\parallel ^{3}} \right|\varphi (r_\parallel) \right\rangle  \   ,
\label{Ebind} 
\end{equation}
where $E_{B}$ is the binding energy for a direct exciton in a kagome lattice monolayer, given by

\begin{eqnarray}
\label{EB0} 
E_{B} = \frac{2  e^{4}/\varepsilon_s^2  \mu}{\hbar^{2}}  \   ,
\end{eqnarray}
and  $\varphi (r_\parallel)$ is the ground state wave function of the 2D direct exciton in a monolayer, given by
\begin{eqnarray}
 \varphi (r_\parallel) = \frac{1}{(2\pi)^{1/2}r_{0}}
e^{- r_\parallel /(2r_{0})} \   .
\label{phi0}
\end{eqnarray}
Therefore, we have  the following expression for dipolar excitons in a double layer kagome lattice  with $D \ll r_{0}$,

\begin{eqnarray}
&& \left\langle \varphi (r) \left| \frac{k
e^{2}D^{2}}{2 \varepsilon r^{3}} \right|\varphi (r) \right\rangle =
\int_{0}^{2\pi} d \varphi \int_{D}^{ \infty} dr\  r\frac{  e^{2}D^{2}}{2
\varepsilon_s r^{3}}
\left|\varphi (r)\right|^{2}  
 \nonumber \\
&=&   \frac{ e^{2}D^{2}}{2 \varepsilon_s r_{0}^{2}}
\int_{D}^{\infty} dr \  \frac{e^{-r/r_{0}}}{r^{2}} =  \frac{
e^{2}D^{2}}{2 \varepsilon_s r_{0}^{3}} \mathrm{I}(D/r_{0}) ,
\label{aver} 
\end{eqnarray}
where the function $\mathrm{I}(D/r_{0})$ is given by
 
\begin{equation}
\mathrm{I}(D/r_{0}) =  \int_{D/r_{0}}^{\infty}dt \frac{e^{-t}}{t^{2}}  . 
\label{ID} 
\end{equation}
Therefore, the exciton binding energy for a dipolar exciton in the double layer
kagome lattice for the case of small $D$, implying that $D \ll r_{0}$, is given by

\begin{equation}
E_{B} = \frac{2  (e^{4}/\varepsilon_s^2)\mu}{\ 
\hbar^{2}} -  
 \frac{ e^{2}D^{2}}{2 \varepsilon_s
r_{0}^{3}}\mathrm{I}(D/r_{0})\   .
\label{E1} 
\end{equation}
The condition $D \ll r_0$ means that the following condition for $c_0$ must be satisfied, i.e., 

\begin{equation}
c_{0}^{2} \ll  \frac{1}{2}  \left(\frac{\hbar\omega}{e^2/\varepsilon_s}  \right) \frac{(\hbar\upsilon_F)^2}{D} 
 \  .
\label{ineq1+} 
\end{equation}

Our numerical results for the exciton binding energy for double layer kagome lattices as a function of the irradiation intensity $c_0$ for large interlayer separation are  presented in Fig.\ \ref{FIG:7}(a). These results are for  frequency about 100  THz   and for  separation D =100 nm . This satisfies the condition $D \gg 8r_0$.  Each curve is  plotted from the minimum value for $c_0$ corresponding to $D=8r_0$.  All plots   show curves increasing  monotonically from its starting value  and tending to $V_0$ as $c_0$ is increased. Figure\ \ref{FIG:7}(b)  compares the exciton binding energy for a pair of closely separated kagome lattices ($D\ll  r_0$) with the binding energy of a monolayer for four chosen values of the hopping parameter $t_h$.  In each case,  the binding energy for a monolayer at low intensity  is  in  good agreement with the double layer result. However,  there is less agreement as $c_0$ is increased.

\begin{figure}
\includegraphics[width=0.6\textwidth]{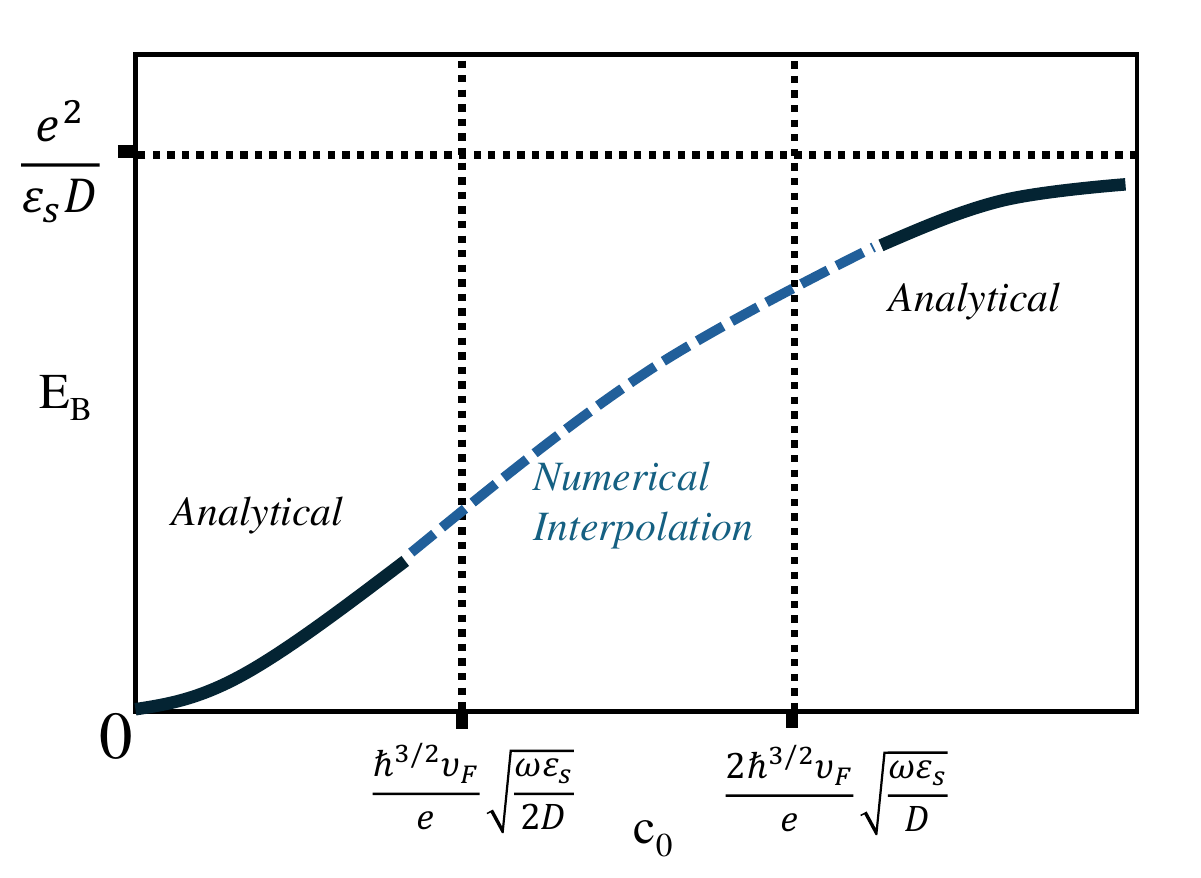}
\caption{ 
 Schematic illustration of the exciton binding energy for kagome lattices  as a function of irradiation intensity $c_0$. The three regions indicate where we have obtained analytical results for small ($D\ll r_0$) and large  ($D\gg r_0$) inter-planar separation as well as the intermediate range where numerical solution of the Schr\"{o}dinger equation  must be executed.}
\label{FIG:8}
\end{figure}

In Fig.\ \ref{FIG:8}, we show schematically the solution of Eq.\ (\ref{eq16}). The figure is divided into three regions. The region on the left is based on our analytical result for the binding energy when $D\ll  r_0$ while the region on the right is based on our analytical results for $D\gg 8r_0$. The center portion of the figure is an interpolation between these two regions. 
\medskip
\par

From our numerical analysis, it is clear that the monolayer excitons are stronger than dipolar excitons for specific values of the irradiation parameters and for fixed kagome lattice. However, the  binding energy of dipolar excitons can also be increased with appropriate tuning of the external field and layer separation.  Although the monolayer excitons have higher binding energy, neither very small  nor very large separation is desirable when you go from monolayer to dipolar excitons. For sufficiently large separation, i.e., for $D\to \infty$,  from  Eqs.\ (\ref{eq17}) and (\ref{eq18}), the Coulomb attraction potential becomes insignificant. Additionally, the large separation gives the condition for $c^2_0$ in Eq.\ (\ref{ineq1}). On the other hand, for the Floquet -Magnus expansion to be valid, we have employed the condition that $c_0\ll \hbar\omega$.  If the inter-planar separation is very small, i.e., $D\to 0$, the term on the right-hand side of the inequality in Eq.~ (\ref{ineq1}) may become very high such that the condition for $c_0$ may not be valid even for high frequency photons in the THz range. Therefore, the separation $D$ should satisfy the condition, $D\gg r_{||}$.   Moreover, the  tuning of the field is associated with the tuning of the energy gap  and the reduced mass of the system and hence  in general the tuning of exciton binding energy.

\section{Collective excitations  for spatially separated layers of electrons and holes} 
\label{sec6}

We determine $E^{\prime}$ and $U(X,Y)$ from Eq.~(\ref{eq4a})  leading to the procedure adopted to solve Eq.~(7) in Ref.~\cite{dipolar4}.  We obtain:
\begin{eqnarray}
U(X,Y) &=& \frac{1}{\sqrt{L_{X}L_{Y}}}
e^{i\left(\frac{P_{X}}{\hbar} - \tau_{X}\right)X}
e^{i\left(\frac{P_{Y}}{\hbar} - \tau_{Y}\right)Y} , \nonumber \\
E^{\prime} &=& \frac{P_{X}^{2}}{2M} +  \frac{P_{Y}^{2}}{2M} -
\frac{\hbar^{2} \tau_{X}^{2}}{2M} - \frac{\hbar^{2}
\tau_{Y}^{2}}{2M} = \frac{P^{2}}{2M} -  \frac{\hbar^{2}
\tau^{2}}{2M}  \   ,
\label{U}
\end{eqnarray}
where $L_XL_Y$ is a normalization area and

\begin{eqnarray}
P^{2} = P_{X}^{2} + P_{Y}^{2}; \hspace{1cm} \tau^{2} =
\tau_{X}^{2} + \tau_{Y}^{2};  \hspace{1cm} \tau_{X} =  \frac{3 M
v_{Fx}}{2\hbar};  \hspace{1cm} \tau_{Y} =  \frac{3 M v_{Fy}}{2\hbar}\   ,
\label{tau} 
\end{eqnarray}
Due to the interlayer separation  $D$, indirect excitons, both in the ground and excited states, have electrical dipole moments. We assume that indirect excitons interact as {\it  parallel}  dipoles.  This is valid when $D$ is larger than the mean separation $\langle  r \rangle $
between the electron and hole along the kagome layers, i.e.,  $D \gg \langle r \rangle$.  Since electrons on a  kagome lattice can be in two valleys, there are four types of excitons in  a kagome lattice. Due to the fact that  all these types of excitons have the same envelope wave functions and energies, we consider below only excitons in one valley. Additionally, we use $n_{0} = n/(4s)$ as the  exciton density in one layer, with
$n$ denoting the total density of exciton, $s$ is the spin
degeneracy  (equal to $4$ for excitons in the double layer of  kagome material).

\medskip
\par
\begin{figure}[h]
\includegraphics[width=0.8\columnwidth]{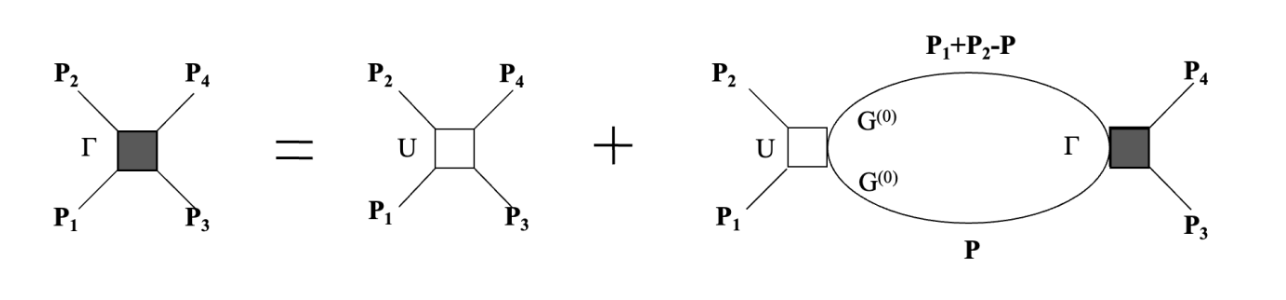}
\caption{ The equation for the vertex $\Gamma$ in the ladder approximation in the momentum representation. }
\label{FIG:9}
\end{figure}

The distinction between excitons and bosons manifests itself in exchange effects \cite{Halperin_Rice,KelKoz}. These effects for excitons with spatially separated electrons and holes in a dilute system with $nr_0^{2} \ll 1$ are suppressed due to the negligible overlap of the wave functions of two excitons  due to the potential barrier, associated with the dipole-dipole repulsion of indirect excitons  (here $r_0 $ is the exciton radius along  the kagome layers).  Two indirect excitons in a dilute system interact as $U(R) = e^{2}D^{2}/(\varepsilon_s R^{3})$, where $R$ is the distance between dipolar excitons along the kagome lattice layers. Little tunneling connected with this barrier is described by the quantity \cite{YuriiL}

\begin{equation} 
\exp\left[- \frac{1}{\hbar }\int_{a}^{r_{0}}dR\  \sqrt{2M
\left(\frac{e^{2}D^{2}}{\varepsilon_s R^{3}} - \frac{\kappa
^{2}}{2M}\right)}\ \right]\   , 
\label{point}
\end{equation}
where

\begin{equation}  
\kappa ^{2} \sim \hbar^{2}\frac{n}{s\ln \left(
s\hbar^{4}\varepsilon_s^{2}/(2\pi n M^2 e^4 D^4) \right)} 
\label{point2}
\end{equation}
\noindent
is the characteristic momentum of the system, $r_{0} =(2Me^{2}D^{2}/\kappa ^{2})^{1/3}$  is the classical turning point for the dipole-dipole interaction.  The small parameter mentioned above has the form $\exp[-2\hbar ^{-1}M^{1/2}eD   r_0^{-1/2}$.  Therefore, at  $T = 0$  K, the dilute gas of excitons, which is a boson system, form a Bose-Einstein  condensate \cite{abrikosov,Griffin}.  Consequently, the system of indirect excitons can be treated by the diagrammatic technique for a bosonic system.  For a dilute 2D  dipolar exciton system (with $n r_0^{2} \ll 1$),   the sum of ladder diagrams is adequate.  The integral equation for vertex $\Gamma $ in the ladder approximation is represented in   Fig.~ \ref{FIG:9}.

\medskip
\par
If we measure energy from  $-\frac{\hbar^{2} \tau^{2}}{2M}$,  where $\tau$ is defined in Eq.\  (\ref{tau} ),  the equation for the vertex has the same form  compared with that in Ref.\  
[\onlinecite{Yudson}] for a 2D boson system in the absence of  anisotropy. We have

\begin{eqnarray}   
\Gamma (\mathbf{p},\mathbf{p}';P) &=& U (\mathbf{p} - \mathbf{p}') +
s \int_{}^{} \frac{d^{2} q}{(2\pi \hbar)^2} \frac{U (\mathbf{p} -
\mathbf{q}) \Gamma (\mathbf{q},\mathbf{p}';P)}{\frac{\kappa ^2}{M}
+\Omega -
\frac{\mathbf{P}^{2}}{4M} - \frac{q^2}{M} + i\delta}, \hspace{0.5cm} \  \  \  \delta  \rightarrow  0^+
 \nonumber\\
\mu_0 &=& \frac{\kappa ^2}{2M} = n_{0}\Gamma _{0} = n_{0} \Gamma (0,0;0)    \   ,
\label{Gamma_Int}
\end{eqnarray}
where  $P = \{\mathbf{P},\Omega \}$, and $\mu_0 $ is the chemical potential of the system.
\medskip
\par
The specific feature of a two-dimensional Bose system is connected with a logarithmic divergence of a two-dimensional scattering amplitude at zero energy \cite{Yudson}. A simple analytical solution of
Eq.~(\ref{Gamma_Int}) for the chemical potential can be obtained if
$\kappa M e^{2} D^2/(\hbar^{3}\varepsilon_s) \ll 1$.  The result for the chemical potential $\mu _0$ is obtained in the form:

\begin{eqnarray}  
\mu_0 =  \frac{\kappa ^2 }{2M} = \frac{\pi \hbar^{2}n}{sM \ln \left[
s\hbar^{4}\varepsilon_s^{2}/\left(2\pi n M^2 e^4 D^4\right) \right]} \  .
\label{Mu}
\end{eqnarray}

The solution of Eq.\   (\ref{Gamma_Int}) for small momentum transfer corresponds to the sound spectrum of collective excitations $\epsilon (P)  = c_s P$  with the sound velocity 

\begin{equation}
c_s = \sqrt{\frac{\Gamma n}{4 sM}} = \sqrt{\frac{\mu_0}{M}}  \  ,
\label{cs}
\end{equation}
 where $\mu_0 $ is defined in Eq.~(\ref{Mu}).  Since excitons have a sound spectrum of collective excitations at small momentum $P$ due to dipole-dipole repulsion, the excitonic superfluidity is possible at low temperatures $T$ in a double layer of kagome lattice because the sound spectrum satisfies the Landau criterion of superfluidity \cite{abrikosov,Griffin}.

\begin{figure} 
\centering
\includegraphics[width=0.45\textwidth]{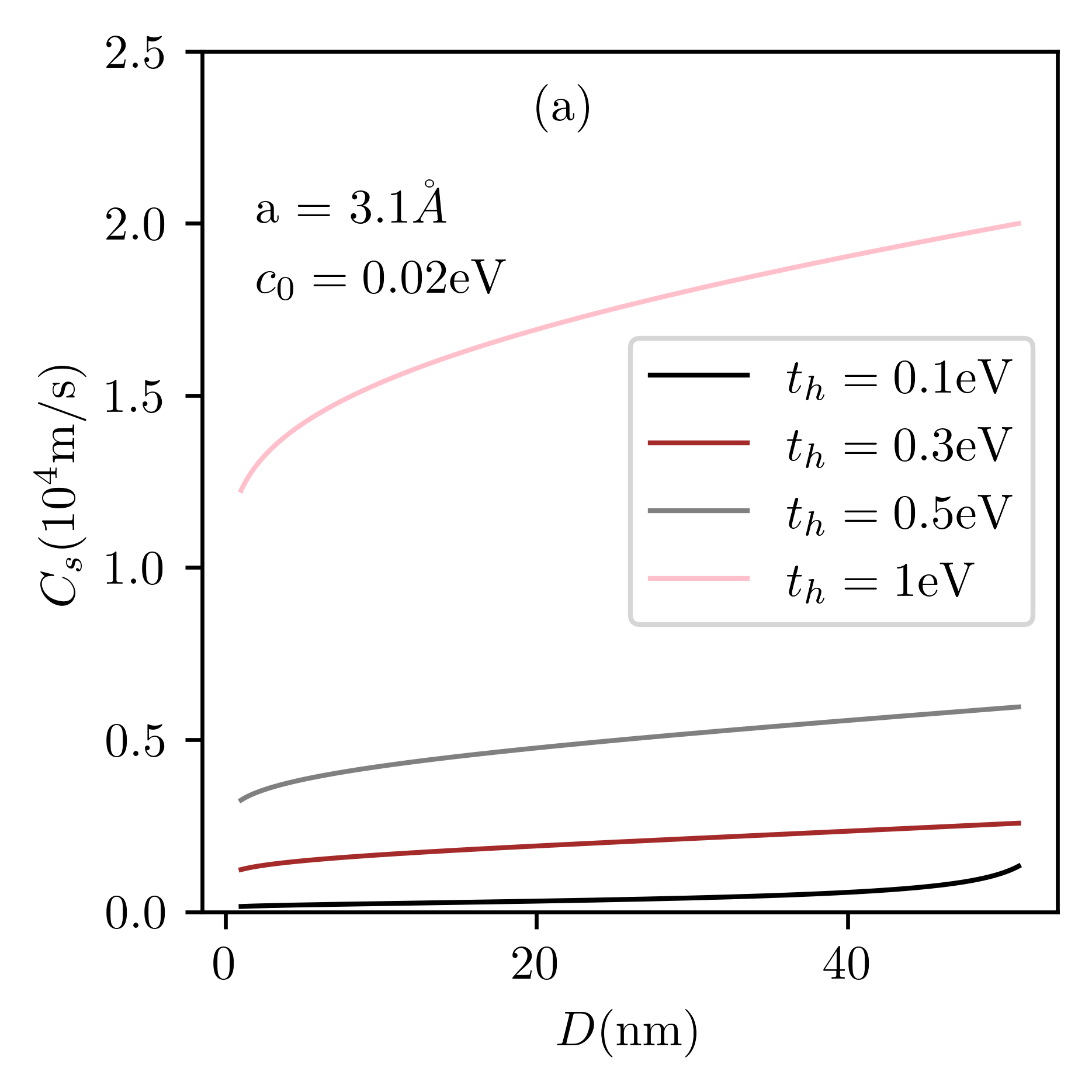}
\includegraphics[width=0.45\textwidth]{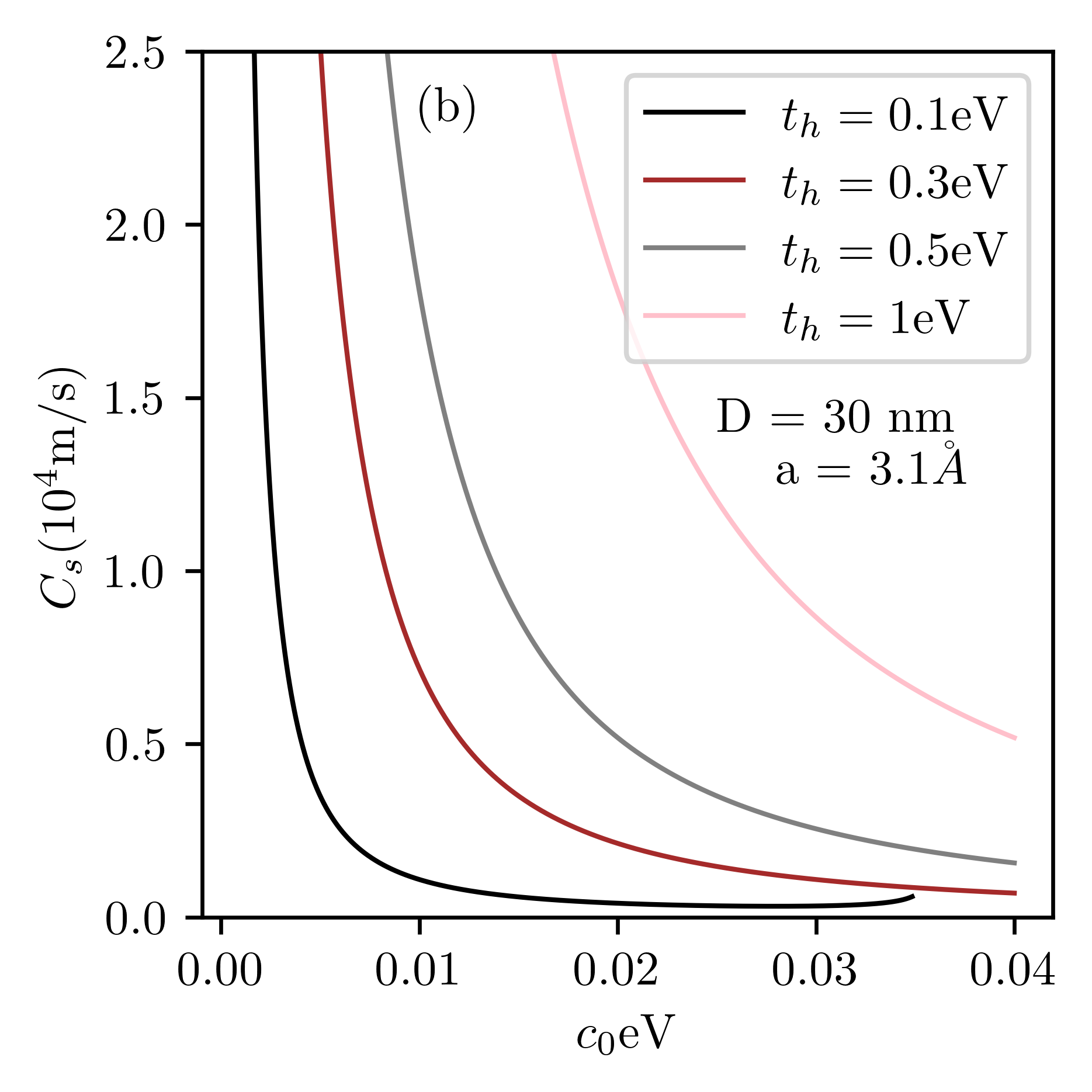}
\caption{(Color online) Speed of sound  for a pair of kagome lattices corresponding to four values of the hopping parameter $t_h$ in (a) as a function of the inter-planar separation. The chosen irradiation intensity is $c_0 = 0.02$ eV. In (b), the sound velocity  is plotted as a function of $c_0$ for inter-layer separation equal to 30 nm. Here, the lattice spacing is assigned $a=3.1\AA$.}
\label{FIG:10}
\end{figure}

\medskip
\par
From the  two equations (\ref{Mu}) and (\ref{cs}), it is clear that the speed of sound for collective excitations of dipolar exciton is a function of the inter-layer separation for a specific kagome lattice and for chosen electric field. As the quantity $D$  appears in the denominator with  the logarithm function, $C_s (D) \to 0$ as $D\to 0^{+}$. However, for large separation, i.e., if $\ln(D^4$) $>$ $\ln(s\hbar^4 \epsilon^2_s / 2\pi n M^2 e^4$), $C_s (D)$ becomes imaginary. Therefore,  the function $C_s (D)$ is real-valued only when $D<(s\hbar^4 \epsilon^2_s / 2\pi n M^2 e^4)^{1/4} $. Hence, within the domain ($0^{+}, (s\hbar^4 \epsilon^2_s / 2\pi n M^2 e^4)^{1/4} $), the speed of sound first increased  with $D$ increased, remains constant and becomes imaginary at $D = (s\hbar^4 \epsilon^2_s / 2\pi n M^2 e^4)^{1/4} $.  Our numerical results are presented in Fig. ~\ref{FIG:10}(a).  The results are for assigned values of $a, c_0, t_h$  and for an EM field of frequency $\sim 100 $ THz. We have chosen h-BN as the dielectric between the layers and the total number density $n=4\times 10^{11} m^{-2}$. 

\medskip
\par
The equations (\ref{Mu}) and (\ref{cs}) also show that the speed of sound does depend on the intensity of irradiation for chosen inter-layer separation. Similar to the dependence on $D$, $C_s(c_0)$ also have a range ($0, (18 (\hbar\omega)^2 t_h^4 a^4 \epsilon^2_s / \pi n D^4 e^4)^{1/4} $), beyond which $C_s(c_0)$ diverges. Within these limits, $C_s$ is decreased as $c_0$ is increased, remains constant and again  increased when $c_0\to (18 (\hbar\omega)^2 t_h^4 a^4 \epsilon^2_s / \pi n D^4 e^4)^{1/4}$. The results for four different values of $t_h$  are exhibited in Fig.~ \ref{FIG:10}(b).

\section{Tunable superfluidity of dipolar excitons in the double layers of kagome lattices.}
\label{sec7}

\begin{figure} 
\centering
\includegraphics[width=0.6\textwidth]{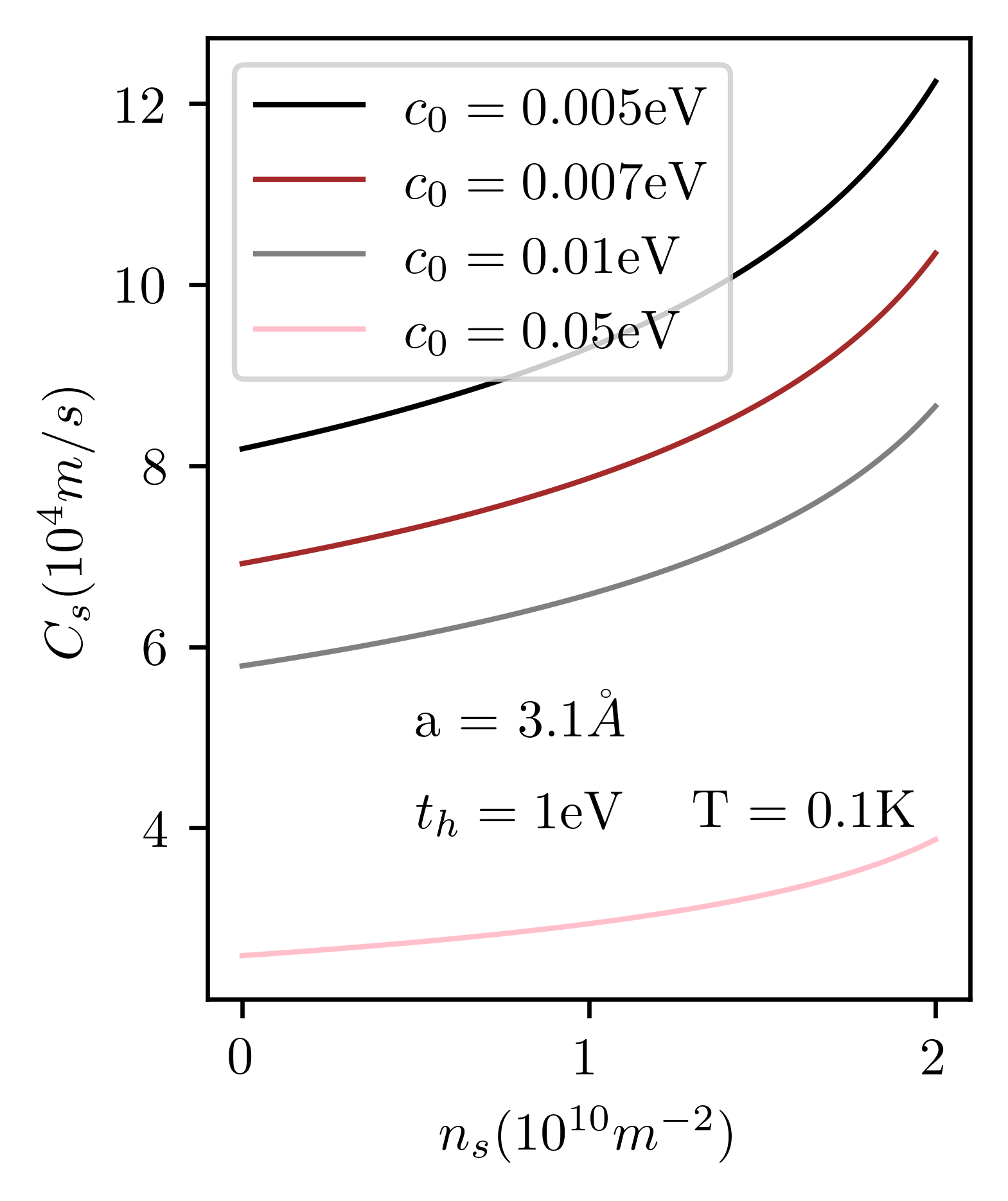}
\caption{(Color online) Speed of sound  for a pair of kagome lattices as a function of the superfluid density $n_s$ at 0.1 K temperature and $t_h = 1.0$ eV for four different values of $c_0$  and  the lattice spacing is chosen as $3.1\AA$.}
\label{FIG:11}
\end{figure}

Dipolar excitons constructed from spatially separated electrons and holes in a double layer of kagome lattice at large interlayer  separations, $D \gg 8r_0$, form a two-dimensional weakly interacting gas of bosons with a pair dipole-dipole repulsion. Therefore,  the  superfluid-normal phase transition for this system is the Kosterlitz-Thouless transition \cite{Kosterlitz, Kosterlitz2} .The temperature of the Kosterlitz-Thouless transition $T_c $ to the superfluid state in a two-dimensional dipolar exciton system is determined by the equation 

\begin{eqnarray}    
T_c = \frac{\pi \hbar ^2 n_s (T_c)}{2 k_B M}\    ,
\label{T_KT}
\end{eqnarray}
where $n_s (T)$ is the superfluid density of the system  of excitons as a function of  temperature $T$, and interlayer distance $D$. Also, $k_B$ is the Boltzmann constant.  The function $n_s (T)$  in Eq.\  (\ref{T_KT}) can be determined from the relation $n_s = n/(4s) - n_n $,   where $n$ is the total density  of the system, and $n_{n}$ is the normal component density.  We determine the normal component density by the usual procedure \cite{abrikosov}.  Suppose the exciton system  moves with  velocity $\mathbf{u}$. At temperatures $T$, dissipating quasiparticles will appear in this system.  Since their density is small at low temperatures, one can assume that the gas of quasiparticles is an ideal Bose gas. To calculate the superfluid component density,  we find the total current of  quasiparticles in a  frame  of reference in which the superfluid component is at rest.  Then,  we obtain the mean total current of two-dimensional  excitons in the coordinate system  moving with velocity ${\bf u}$:

\begin{equation}
\left\langle \mathbf{J} \right\rangle = \frac{1}{M}
\left\langle \mathbf{P} \right\rangle =\frac{s}{M} \int_{}^{}
\frac{d\mathbf{P}}{(2\pi \hbar)^{2}}\   \mathbf{P} f\left[\epsilon(P) - \mathbf{P}\cdot \mathbf{u} \right]\    ,
\label{nnor}    
\end{equation}
where $f\left[\epsilon (P)\right] = \left(\exp\left[\varepsilon
(P)/(k_{B}T)\right] - 1\right)^{-1}$ is the equilibrium Bose-Einstein distribution function. Expanding the expression within the integral to first order by $\mathbf{P}\cdot \mathbf{u}/(k_{B}T)$, we have:

\begin{eqnarray}  
\langle \mathbf{J} \rangle = - s\frac{\mathbf{u}}{2
M}\int\frac{d\mathbf{P}}{(2\pi \hbar)^{2}}P^{2}\frac{\partial
f\left[\epsilon (P)\right]}{\partial \epsilon}=
  \frac{3 \zeta (3)s }{2 \pi \hbar^{2}}\frac{k_{B}^{3}T^3}{M c_s^4} \mathbf{u}\  ,
\label{J_Tot}
\end{eqnarray}
where $\zeta (z)$ is the Riemann zeta function ($\zeta (3) \simeq  1.202$). Then, we define the normal component density $n_{n}$ as  \cite{abrikosov}

\begin{eqnarray}   
\langle  \mathbf{J} \rangle = n_n  \mathbf{u}\   .
\label{J_M}
\end{eqnarray}
Comparing  Eqs.~ (\ref{J_Tot}) and (\ref{J_M}),  we obtain an expression for the normal density $n_{n}$. Consequently, we have for the superfluid density:

\begin{eqnarray}
n_s = n/(4s) - n_n = n/(4s) -
 \frac{3 \zeta (3) }{2 \pi \hbar^{2}} \frac{k_{B}^{3}T^3}{c_s^4 M}\  
\label{n_s} 
\end{eqnarray}
which yields

\begin{equation}
c_{s}=\left(s\frac{3s\zeta(3)k_{B}^{3}T^{3}}{2\pi\hbar^{2}M}\times \frac{1}{\frac{n}{4s} -n_{s}} \right)^{1/4}\   
\label{csns}
\end{equation}
\noindent
for the speed of sound as a function of superfluid density $n_s$. This expression for $C_{s}$ does not depend on the inter-plane separation $D$ and constrains the superfluid density to be less than one fourth of total number density otherwise the speed becomes imaginary.  Substituting into Eq.~ (\ref{csns}) the mass for the system $M=2m_{h}$=$\frac{2c_{0}^{2}}{\hbar\omega\upsilon_{F}^{2}} $ we have  plotted  $C_{s}$ as a function of $n_s$  for chosen temperature   T=$0.1$ K,  n=$4\times 10^{11}m^{-2}$ and four different values of $c_0$ in Fig.\  \ref{FIG:11}. Other parameters, such as $a$ and $t_h$, are also given on the plot.These results show that $C_s$ is increased monotonically with $n_s$  as for $n/4 > n_s$, the term $(n/4 - n_s)$ is decreased in the denominator when the  superfluid density is increased. Consequently, $C_s$ is  increased. This behavior is consistent with the dependence of $C_s$ on $c_0\propto E_0$ depicted   in Fig.\  ~ \ref{FIG:10}.  When $n_s$ is increased from zero and becomes equal to n/4, $C_s(n_s)$ diverges. This is why it is not possible for the speed of sound to be infinite for fixed exciton density. Here, the superfluid density can be tuned by the irradiation parameter at fixed low temperature.

\begin{figure}[!h]    
\includegraphics[width=0.45\textwidth]{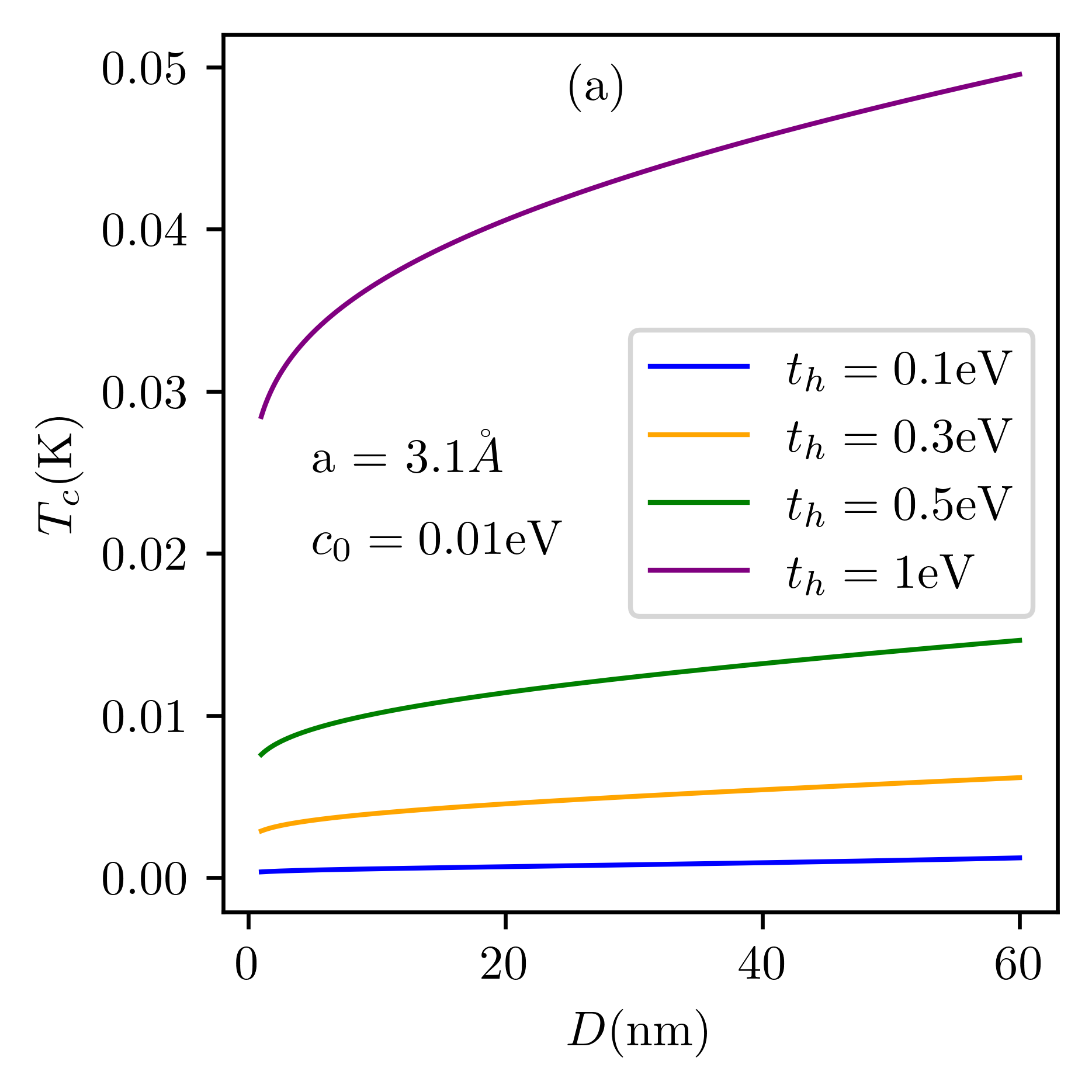}
\includegraphics[width=0.45\textwidth]{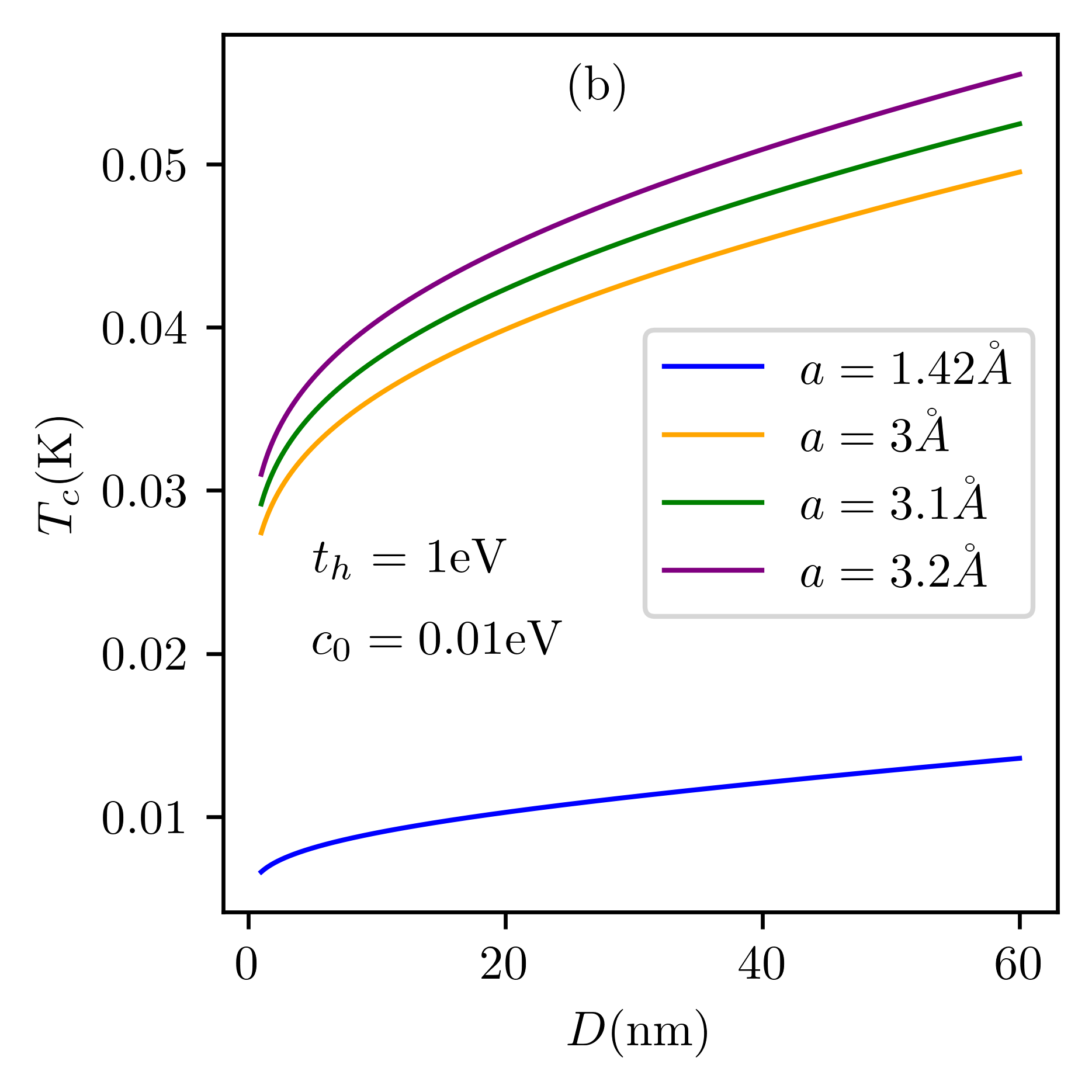}
\caption{(Color online) Kosterlitz-Thouless transition temperature as a function of inter-planer separation. The curves are for four different values of $t_h$ for fixed lattice spacing $a=3.1\AA$ in (a) and for four different lattice spacings with $t_h=1.0$ eV in (b). The intensity of the external irradiation $c_0 = 0.01$ eV which is less less than the photon energy $\hbar\omega =0.1$ eV. The graphs are plotted using Eq.~\ref{tct} }
\label{FIG:12} 
\end{figure}

\medskip
\par
In a two-dimensional  system, superfluidity of dipolar excitons appears below the Kosterlitz-Thouless transition temperature   given in Eq.~(\ref{T_KT}), where only coupled vortices are present  \cite{Kosterlitz}.  Employing Eq.~(\ref{n_s}) for the density $n_{s}$ of the superfluid component, we obtain an equation for the Kosterlitz-Thouless transition temperature $T_{c}$.  Its solution is

\begin{equation}
T_c = \left[\left( 1 + \sqrt{\frac{32}{27}\left(\frac{s
Mk_{B}T_{c}^{0}}{\pi \hbar^{2} n}\right)^{3} + 1} \right)^{1/3} -
\left( \sqrt{\frac{32}{27} \left(\frac{s Mk_{B}T_{c}^{0}}{\pi
\hbar^{2} n}\right)^{3} + 1} - 1 \right)^{1/3}\right]
\frac{T_{c}^{0}}{ 2^{1/3}} \  .
\label{tct}  
\end{equation}
In this notation,  $T_{c}^{0}$ is an auxiliary quantity, equal to the temperature at which the superfluid density vanishes in the mean-field approximation (i.e., $n_{s}(T_{c}^{0}) = 0$),

\begin{equation}
   T_c^0 = \frac{1}{k_{B}} \left( \frac{ \pi \hbar^{2} n
c_s^4 M}{6 s\zeta (3)} \right)^{1/3} \  .
\label{tct0} 
\end{equation}
The temperature $T_{c}^{0} = T_{c}^{0}(B,D)$ may be used as a rough  estimate of the crossover region where local superfluid density appearers for dilute dipolar exciton system on the scales smaller or of an order of mean intervortex separation in the system. The local
superfluid density can manifest itself in local optical properties  or local transport properties.

\medskip
\par
According to Eqs.\  (\ref{tct}) and~(\ref{tct0}), the temperature $T_c$ for the onset of superfluidity due to the  Kosterlitz-Thouless transition depends on the dipolar exciton density, interlayer separation   and  the reduced mass of the exciton.  As the mass depends on the external field intensity, for assigned  $c_0 , \omega$. For fixed dipolar exciton density,  the temperature increases as a function of inter-layer separation $D$. This is also shown in Fig.~\ref{FIG:12} for chosen values of  material parameter $a$ and $t_h$. We have used $c_0 = 0.01 eV < \hbar \omega = 0.1$ eV.

\section{Conclusions and Discussion}
\label{sec8}

The appearance of local superfluid density above $T_{c}$ can be manifested, for example, in observations of the  temperature dependence of the exciton diffusion on intermediate distances by performing local measurements of exciton photoluminescence at two points using optical fibers or pinholes. The superfluid state at $T < T_{c}$ can manifest itself in the existence of persistent oppositely directed electric currents in
each kagome lattice layer.  The interlayer resistance relating to the drag of electrons and holes can also be a sensitive indicator of the transition to the superfluid of the electron-hole
system~\cite{Vignale}.

\medskip
\par
In conclusion, we have systematically calculated the binding energies  of excitons and exciton energies in monolayer and double layer kagome lattices in the presence of circularly polarized irradiation. We explored the possibilities of tuning  the phase transition in kagome lattice from the semiconducting phase to the EI phase by applying circularly polarized radiation.   Since conductive properties of the semiconductor phase and EI phase are different, kagome lattices can be applied for developing novel switches. We also studied BEC and superfluidity of dipolar excitons in two layers of kagome lattice with an applied external voltage in the presence of pumping by circularly polarized light.  

\medskip 
\par
The superfluid density $n_{s}(T)$ and the temperature $T_{c}$ of the Kosterlitz-Thouless phase transition to the superfluid state have been calculated as functions of the parameters  pertaining to the circularly polarized  irradiation light. We have shown that at chosen exciton density $n$, the Kosterlitz-Thouless temperature $T_{c}$ for the onset of superfluidity of dipolar excitons  increases when the density $n$ and  the interlayer separation are increased. We propose that the exciton binding energies, the spectrum of collective excitations, the superfluid density, and critical temperature of superfluidity can be tuned by applying circularly polarized irradiation. Since the exciton binding energy is an increasing function of the parameter of the circular polarized radiation $c_0$, while the sound velocity and the critical temperature of the superfluidity are decreasing functions of $c_0$, one has to find optimal $c_0$ for the corresponding interlayer separation $D$ to achieve large enough exciton binding energy and $T_c$. Since the band gap and the exciton binding energy depend on  irradiation intensity, we propose that one can tune the phase transition in kagome lattice from the semiconducting phase to the EI phase by applying the circularly polarized irradiation.

\begin{acknowledgements}
G.G. was supported by Grant No. FA9453-21-1-0046 from the Air Force Research Laboratory (AFRL). 
 G.G. gratefully acknowledges funding from the U.S. National Aeronautics and Space Administration (NASA) via the NASA-Hunter College Center for Advanced Energy Storage for Space  under cooperative agreement 80NSSC24M0177.
\end{acknowledgements}

{}


\begin{thebibliography}{}


\bibitem{KS} Christos Kourris and Matthias Vojta,  Kondo screening and coherence in kagome local-moment metals: Energy scales of heavy fermions in the presence of flat bands, Phys. Rev. B {\bf 108}, 235106   (2023).

\bibitem{K1}   Y. Tokiwa, C. Stingl, M. S. Kim, T. Takabatake, and P. Gegenwart, Characteristic signatures of quantum criticality driven by geometrical frustration, Sci. Adv. {\bf 1}, e1500001 (2015).

\bibitem{K2} S. Kittaka, Y. Kono, S. Tsuda, T. Takabatake, and T. Sakakibara, Field-Angle-Resolved Landscape of Non-Fermi-Liquid Behavior in the Quasi-Kagome Kondo Lattice CeRhSn, 
J. Phys. Soc. Jpn. {\bf 90}, 064703 (2021).

\bibitem{K3} V. Fritsch, N. Bagrets, G. Goll,W. Kittler, M. J.Wolf, K. Grube, and C.-L. Huang, and H. von Löhneysen, Approaching quantum criticality in a partially geometrically frustrated heavy-fermion metal, Phys. Rev. B {\bf 89},054416 (2014).

\bibitem{K4}   S. Lucas, K. Grube, C.-L. Huang, A. Sakai, S. Wunderlich, E. L. Green, J. Wosnitza, V. Fritsch, P. Gegenwart, O. Stockert, and H. v. Löhneysen, Entropy Evolution in the Magnetic Phases of Partially Frustrated CePdAl,  Phys. Rev. Lett. {\bf 118}, 107204 (2017).

\bibitem{K5}   H. Zhang, J. Zhao, M. Lv, S. Hu, Y. Isikawa, Y. Yang, Q. Si, F. Steglich, and P. Sun, Quantum-critical phase from frustrated magnetism in a strongly correlated metal, Nat. Phys. {\bf 15}, 1261 (2019).

 \bibitem{K6}   P. G. Niklowitz, G. Knebel, J. Flouquet, S. L. Bud’ko, and P. C. Canfield,   Field-induced non-Fermi-liquid resistivity of stoichiometric YbAgGe single crystals, Phys. Rev. B {\bf 73}, 125101 (2006).

\bibitem{K7}   Y. Tokiwa, M. Garst, P. Gegenwart, S. L. Bud’ko, and P. C. Canfield, Quantum Bicriticality in the Heavy-Fermion Metamagnet YbAgGe  Phys. Rev. Lett. {\bf 111}, 116401 (2013).

\bibitem{K8}  W. Xie, F. Du, X. Y. Zheng, H. Su, Z. Y. Nie, B. Q. Liu, Y. H. Xia, T. Shang, C. Cao, M. Smidman, T. Takabatake, and H. Q. Yuan, Semimetallic Kondo lattice behavior in YbPdAs with a distorted kagome structure, Phys. Rev. B {\bf 106}, 075132 (2022).

\bibitem{K9}   S. Nakatsuji, Y. Machida, Y. Maeno, T. Tayama, T. Sakakibara, J. van Duijn, L. Balicas, J. N.Millican, R. T.Macaluso, and J. Y. Chan, Metallic Spin-Liquid Behavior of the Geometrically Frustrated Kondo Lattice  Pr2Ir2O7,   Phys. Rev. Lett. {\bf 96}, 087204 (2006).

\bibitem{K10}   Y. Machida, S. Nakatsuji, S. Onoda, T. Tayama, and T. Sakakibara, Nature (London) {\bf 463}, 210 (2010).

\bibitem{K11}   Y. Tokiwa, J. J. Ishikawa, S. Nakatsuji, and P. Gegenwart, Quantum criticality in a metallic spin liquid, Nat. Mater. {\bf 13}, 356 (2014).

\bibitem{K12}   L. Ye, S. Fang, M. Gu Kang, J. Kaufmann, Y. Lee, J. Denlinger, C. Jozwiak, A. Bostwick, E. Rotenberg, E. Kaxiras, D. C. Bell, O. Janson, R. Comin, and J. G. Checkelsky,  Hopping frustration-induced flat band and strange metallicity in a kagome metal, Nature Physics {\bf 20}, 610-614 (2024).

\bibitem{K13}    L. Huang and H. Lu, Protracted Kondo screening and kagome bands in the heavy-fermion metal Ce$_3$Al, Phys. Rev. B {\bf 102}, 155140 (2020).

\bibitem{Naf2} Man Li, Qi Wang, Guangwei Wang, Zhihong Yuan, Wenhua Song, Rui Lou, Zhengtai Liu, Yaobo Huang, Zhonghao Liu, Hechang Lei, Zhiping Yin, and Shancai Wang, Dirac cone, flat band, and saddle point in kagome magnet YMn$_6$Sn$_6$,
Nature Communications {\bf 12}, 3129 (2021).

\bibitem{Naf1} Luting Xu and Fan Yang,
Band structures of strained kagome lattices,  Chin. Phys. B {\bf 33}, 027101 (2024).

\bibitem{lacroix2011} C. Lacroix, P. Mendels, and F. Mila, Introduction to frustrated magnetism: materials, experiments, theory, vol. {\bf 164} (Springer Science \& Business Media, 2011).

\bibitem{lee2007} S.-H. Lee, H. Kikuchi, Y. Qiu, B. Lake, Q. Huang,
K. Habicht, and K. Kiefer, Quantum-spin-liquid states in the two-dimensional kagome antiferromagnets ZnxCu4-x(OD)6Cl2, Nature Materials {\bf 6}, 853 (2007).

\bibitem{balents2010} L. Balents, Spin liquids in frustrated magnets, 
  Nature {\bf 464}, 199 (2010).

\bibitem{yan2011} S. Yan, D. A. Huse, and S. R. White, Spin-Liquid Ground State of the S = 1/2 Kagome Heisenberg Antiferromagnet, Science {\bf 332}, 1173 (2011).

\bibitem{jiang2012} H.-C. Jiang, Z. Wang, and L. Balents, Identifying topological order by entanglement entropy, Nature Physics {\bf 8}, 902 (2012).

\bibitem{han2012} T.-H. Han, J. S. Helton, S. Chu, D. G. Nocera, J. A.
Rodriguez-Rivera, C. Broholm, and Y. S. Lee, Fractionalized excitations in the spin-liquid state of a kagome-lattice antiferromagnet, Nature {\bf 492}, 406 (2012).

\bibitem{norman2016} M. Norman, Wrinkles, creases, and cusps in growing soft matter, Reviews of Modern Physics {\bf 88}, 041002 (2016).

\bibitem{liao2017} H.-J. Liao, Z.-Y. Xie, J. Chen, Z.-Y. Liu, H.-D. Xie, R.-
Z. Huang, B. Normand, and T. Xiang, Gapless Spin-Liquid Ground State in the s = 1/2 kagome antiferromagnet,  Physical Review Letters {\bf 118}, 137202 (2017).

\bibitem{zhou2017} Y. Zhou, K. Kanoda, and T.-K. Ng, Quantum spin liquid states,
Reviews of Modern Physics {\bf 89}, 025003 (2017).

\bibitem{lauchli2019} A. M. Lauchli, J. Sudan, and R. Moessner, S=1/2 kagome Heisenberg antiferromagnet revisited
Physical Review B {\bf 100}, 155142 (2019).

\bibitem{bergman2008} D. L. Bergman, C. Wu, and L. Balents, Band touching from real-space topology in frustrated hopping models,  Physical Review
B {\bf 78}, 125104 (2008).

\bibitem{yang2014} B.-J. Yang and N. Nagaosa, Emergent Topological Phenomena in Thin Films of Pyrochlore Iridates, 
Physical Review Letters {\bf 112}, 246402 (2014).

\bibitem{mielke1991} A. Mielke, Ferromagnetism in the Hubbard model on line graphs and further considerations, 
Journal of Physics A: Mathematical and General {\bf 24}, 3311 (1991).

\bibitem{mielke1992} A. Mielke, Exact ground states for the Hubbard model on the kagome lattice Journal of Physics A: Mathematical and General {\bf 25}, 4335 (1992).

\bibitem{bilitewski2018} T. Bilitewski and R. Moessner, Disordered flat bands on the kagome lattice, Physical Review B {\bf 98}, 235109 (2018).


\bibitem{mallah} C. Barreteau, F. Ducastelle, and T. A bird's eye view on the flat and conic band world of the honeycomb and kagome lattices: towards an understanding of 2D metal-organic frameworks electronic structure,
J. Phys.: Condens. Matter {\bf 29}, 465302 (2017).

\bibitem{Mazin2014}I. I. Mazin, H. O. Jeschke, F. Lechermann, H. Lee, M. Fink, R. Thomale and R. Valent\'i,  Theoretical prediction of a strongly correlated Dirac metal,  Nature Comm. \textbf{5}, 4261 (2014).

\bibitem{Ye2018}L. Ye, M. Kang, J. Liu, F. von Cube, C. R. Wicker, T. Suzuki, C. Jozwiak, A. Bostwick, E. Rotenberg, D. C. Bell, L. Fu, R. Comin and J. G. Checkelsky,  Massive Dirac fermions in a ferromagnetic kagome metal, Nature \textbf{555}, 638 (2018).

\bibitem{Leykam2018}D. Leykam, A. Andreanov and S. Flach, Artificial flat band systems: from lattice models to experiments, Adv. Phys.: X \textbf{3}, 1473052 (2018).

\bibitem{nature18}  E. Tang,  J.-W.Mei, and X.-G. Wen, High-temperature fractional quantum Hall states. Phys. Rev. Lett. {\bf 106}, 236802 (2011).

\bibitem{nature19}  K. Sun, Z. Gu,  H. Katsura, and S. Das Sarma,  Nearly flatbands with nontrivial topology. Phys. Rev. Lett. {\bf 106}, 236803 (2011).

\bibitem{nature20} T. Neupert, L. Santos,  C. Chamon, and C. Mudry, Fractional quantum Hall states at zero magnetic field. Phys. Rev. Lett. {\bf 106}, 236804 (2011).

\bibitem{nature21} Y.  Cao, et al. Correlated insulator behavior at half-filling in magic-angle graphene superlattices. Nature {\bf 556}, 80–84 (2018).

\bibitem{nature22}  Y. Cao, et al., Unconventional superconductivity in magic-angle graphene superlattices. Nature {\bf 556}, 43–50 (2018).


\bibitem{para1} Y. E. Lozovik, and V. I. Yudson. Feasibility of superfluidity of paired spatially separated electrons and holes; a new superconductivity mechanism. \textit{JETP}  \textbf{22}.11 (1975).

\bibitem{para2} Y. E. Lozovik, and V. I. Yudson. A new mechanism for superconductivity: pairing between spatially separated electrons and holes. \textit{JETP} \textbf{71}  738 (1976).

\bibitem{littlewood} X. Zhu, \textit{et al.} Exciton condensate in semiconductor quantum well structures \prl \textbf{74}  1633 (1995).

\bibitem{para4} D. W. Snoke. Spontaneous Bose coherence of excitons and polaritons. \textit{Science} \textbf{298}  1368-1372 (2002).

\bibitem{para5} L.V. Butov. Condensation and pattern formation in cold exciton gases in coupled quantum wells. \textit{Journ. of Phys.: Cond. Matt.} \textbf{16}.50  R1577 (2004).

\bibitem{para6} J. P. Eisenstein,  and A. H. MacDonald. Bose–Einstein condensation of excitons in bilayer electron systems. \textit{Nature} \textbf{432}  691-694 (2004).

\bibitem{para7} D. W. Snoke In \textit{Quantum Gases: Finite Temperature and Non-equilibrium Dynamics; Cold Atom Series}; N. P. Proukakis, A. A. Gardiner,M. J. Davis, and M. H. Szymanska, Eds.; Imperial College Press: London, UK, 2013; Volume 1, p. 419.

\bibitem{para8} S. Saberi-Pouya, \textit{et al.} Experimental conditions for the observation of electron-hole superfluidity in GaAs heterostructures. \prb \textbf{101}  140501 (2020).


\bibitem{para16} F. C. Wu, F. Xue, and A. H. MacDonald. Theory of two-dimensional spatially indirect equilibrium exciton condensates. \prb \textbf{92}  165121 (2015).

 \bibitem{para19} S. Conti, \textit{ et al.} Doping-dependent switch from one-to two-component superfluidity in coupled electron-hole van der Waals heterostructures. \prb \textbf{101}  220504 (2020).


\bibitem{para17} O. L. Berman, and R. Ya. Kezerashvili. High-temperature superfluidity of the two-component Bose gas in a transition metal dichalcogenide bilayer. \prb \textbf{93}  245410 (2016).

\bibitem{dipolar1} O. L. Berman, and R. Ya.  Kezerashvili. Superfluidity of dipolar excitons in a transition metal dichalcogenide double layer \prb \textbf{96}  094502 (2017).


\bibitem{dipolar2} O. L. Berman, G Gumbs, and R. Ya. Kezerashvili. Bose-Einstein condensation and superfluidity of dipolar excitons in a phosphorene double layer \prb \textbf{96}  014505 (2017).


\bibitem{dipolar3} O. L. Berman, G. Gumbs, G. P. Martins, and P. Fekete. Superfluidity of Dipolar Excitons in a Double Layer of $\alpha$ - $T_3$ with a Mass Term \textit{Nanomaterials} \textbf{12}.9  1437 (2022).

\bibitem{dipolar4} A. N. Arafat, O. L. Berman, and G. Gumbs. Superfluidity of indirect momentum space dark dipolar excitons in a double layer with anisotropic tilted semi-Dirac bands \prb \textbf{109} 224506 (2024).

\bibitem{abrikosov} A.  A.  Abrikosov, L. P. Gorkov, I. Y. Dzyaloshinskii. \textit{Quantum Field Theoretical Methods in Statistical Physics} (Prentice-Hall, Englewood Cliffs, NJ, 1963).


\bibitem{landau} V. Ginzburg,  and L. Landau. On superconductivity and superfluidity \textit{Physik Journal} \textbf{8}.11  56 (2009).


\bibitem{pitbook} L. Pitaevskii, and S. Stringari. \textit{Bose-Einstein condensation and superfluidity}.  Oxford University Press, 2016.



\bibitem{expsec2} J. Cutshall \textit{et al.} Imaging interlayer exciton superfluidity in a 2D semiconductor heterostructure. \textit{Sci. Adv.} \textbf{11}.1 (2025).



\bibitem{Eugene} Eugene Kogan, and Godfrey Gumbs, Green's functions and DOS for some 2D lattices, Graphene,{\bf 10}, No.1, December 2 (2021).

 \bibitem{ciola} R. Ciola, K. Pongsangangan, R. Thomale and L. Fritz, Chiral symmetry breaking through spontaneous dimerization in kagome metals, Phys. Rev. B {\bf 104}, 245138 (2021).

\bibitem{AJP} B. Zaslow and Melvin E. Zandler, Two-Dimensional Analog to the Hydrogen Atom, Am. J. Phys. {\bf 35}, 1118–1119 (1967). 
 
 \bibitem{arora_1} A. Arora, K. Nogajewski, M. Molas, M. Koperski and M. Potemski,  Exciton band structure in layered MoSe$_2$: from a monolayer to the bulk limit, Nanoscale, {\bf 7}, 20769 (2015).

 \bibitem{arora_2} A. Arora, M. Koperski,  K. Nogajewski, J. Marcus and C. Faugeras,  Excitonic resonances in thin films of  WSe$_2$: from  monolayer to bulk material, Nanoscale, {\bf 7}, 10421 (2015).
 
 \bibitem{Nishanov} Yu.~E. Lozovik and V.~N. Nishanov,  Wannier-Mott excitons in layer structures  and near an interface  of two mediam, Sov.~Phys.~Solid~State {\bf 18}, 1905  (1976).
 
 
 \bibitem{Halperin_Rice} B.~I. Halperin and T.~M. Rice,  The Excitonic State at the Semiconductor-Semimetal Transition, Solid State Phys. {\bf  21}, 115 (1968).


\bibitem{KelKoz} L.~V. Keldysh and A.~N. Kozlov, Collective properties of excitons in semiconductors, JETP {\bf 27}, 521 (1968).

\bibitem{YuriiL} Oleg L. Berman, Yurii E. Lozovik, and Godfrey Gumbs, Bose-Einstein condensation and superfluidity of magnetoexcitons in bilayer graphene, Phys. Rev. B  {\bf 77}, 155433   (2008).

\bibitem{Griffin} A. Griffin, {\em Excitations in a Bose-Condensed Liquid} (Cambridge University Press, Cambridge, England, 1993).

\bibitem{Yudson} Yu.~E. Lozovik and V.~I. Yudson,  On the ground state of the two-dimensional non-ideal bose gas,  Physica  {\bf A 93}, 493 (1978).

\bibitem{Kosterlitz} J. M. Kosterlitz and D. J. Thouless,  Ordering, metastability and phase transitions in two-dimensional systems,  J.~Phys. {\bf C 6},  1181 (1973).
\bibitem{Kosterlitz2}  D.~R.Nelson and J.~M. Kosterlitz,  Universal Jump in the Superfluid Density of Two-Dimensional Superfluids,    Phys. Rev. Lett.  {\bf 39}, 1201 (1977).

\bibitem{Vignale} G. Vignale and A. H. MacDonaldand, Drag in Paired Electron-Hole Layers, Phys. Rev. Lett. {\bf 76} 2786 (1996).
 
 
 
 
 
\end{thebibliography}
\end{document}